\documentclass[a4paper,twocolumn,10pt]{article}
\usepackage{geometry}
\usepackage[backend=biber,style = chem-acs,doi=true,maxnames=10,minnames=10,articletitle=true]{biblatex}
\usepackage{graphicx}

\usepackage{siunitx}
\DeclareSIUnit{\sq}{sq} 

\usepackage{amsmath}
\usepackage{subcaption}

\usepackage{hyperref}
\usepackage{cleveref}
\usepackage{float}       
\usepackage{mathtools}
\usepackage{csquotes}    
\usepackage{tabularx}    



\usepackage{authblk}
\author[1]{Constantin Bach}
\affil[1]{Institut für Physik, Technische Universität Chemnitz, 09126 Chemnitz, Germany}
\author[2,3]{Dong Won Kim}
\author[2,3]{Yitian Du}
\author[2,3]{Yana Vaynzof}
\affil[2]{Leibniz-Institut für Festkörper- und Werkstoffforschung Dresden (IFW), 01069 Dresden, Germany}
\affil[3]{Chair for Emerging Electronic Technologies, Technische Universität Dresden, 01187 Dresden, Germany}
\author[1]{Carsten Deibel*}

\title{Operando Charge Carrier Dynamics by Intensity-Modulated Photoluminescence: From Perovskite Thin Films to Solar Cells}
\date{*Email: deibel@physik.tu-chemnitz.de}

\begin{document}

\maketitle
\begin{abstract}
 The charge-carrier lifetime in halide perovskites often varies by orders of magnitude with the injection level, complicating the analysis of conventional time-resolved photoluminescence and comparisons of reported lifetimes. We use intensity-modulated photoluminescence spectroscopy (IMPLS) as a frequency-domain alternative. Carried out under operating conditions, it yields the lifetime at a specific injection level, e.g., one-sun-equivalent illumination. On thin films, IMPLS and steady-state photoluminescence yield comparable lifetimes. For perovskite/transport-layer stacks, the IMPLS response becomes more complex. A dedicated transfer function disentangles charge transfer from interface recombination and predicts an upper limit on the charge-carrier lifetime in the solar cell. In devices, electrical methods such as intensity-modulated photovoltage spectroscopy (IMVS) often fail to measure the lifetime due to capacitive effects. IMPLS shows not only the same two characteristic frequencies as IMVS, but also a third one, thereby resolving the lifetime itself.
\end{abstract}

\begin{refsection} 
    
\clearpage
The charge-carrier lifetime is an essential property of the absorber material for the efficiency of a solar cell~\cite{Kaienburg2020HowSolar,Beck1996MobilityLifetime,Coutinho2015InfluenceCharge}. It can be determined using many different techniques, e.g., based on photoluminescence (PL), photoconductivity or photovoltage. Within the field of halide perovskite solar cells, time-resolved PL (TRPL) is one of the most widely used methods. As the charge-carrier lifetime in these materials often strongly depends on the injection level and can vary by orders of magnitude for the same sample~\cite{Yuan2025UnderstandingPowerLaw,Aalbers2026LowTemperatures,Tiede2026ComparativeAnalysis,Levine2018CanWe}, relating the decay time extracted from a transient experiment to a specific operating point of the solar cell is not straightforward~\cite{Kruckemeier2021UnderstandingTransient,Kruckemeier2021ConsistentInterpretation}. Moreover, shallow traps in perovskites~\cite{Yuan2024ShallowDefects,Li2023ShallowTrapsinduced,Marunchenko2024HiddenPhotoexcitations} may influence recombination dynamics differently under transient and steady-state conditions, so that the apparent lifetime depends on the measurement condition.

The frequency-domain counterpart of TRPL, intensity-modulated photoluminescence spectroscopy (IMPLS), dates back to 1926~\cite{Gaviola1926AbklingungszeitenFluoreszenz} and has had numerous applications since then~\cite{Bruggemann2006ModulatedPhotoluminescence, Giesecke2010MinorityCarrier,Reklaitis2016PhotoluminescenceDecay,King2024FrequencydomainTimeresolved}. In analogy to similar methods such as intensity-modulated photovoltage spectroscopy (IMVS), the sample is excited with sinusoidally modulated light and the signal response -- in this case, the PL -- is evaluated in terms of its phase shift and amplitude relative to the excitation, which we write in the corresponding complex notation throughout. As a small-signal technique operating around a working point (e.g., one-sun-equivalent illumination), IMPLS bypasses the aforementioned challenges of TRPL.

IMPLS has so far been rarely applied to perovskites. The few existing reports focused either on low frequencies, investigating the dynamics of mobile ions rather than charge carriers~\cite{Gillespie2025IntensityModulatedPhotoluminescence,Gillespie2026PhotoluminescenceMapping,Gillespie2026ResolvingMobile}, or used a single frequency in the kHz range~\cite{Mahaffey2023MeasuringAbsorber}, which risks significantly overestimating the charge-carrier lifetime~\cite{Moron2021AnalyticalModel,Berenguier2025ContactlessDefects}. Instead, we show IMPLS data spanning orders of magnitude in both frequency (up to tens of MHz) and illumination levels, enabling us to study charge-carrier dynamics in perovskite thin films, absorber/transport layer (TL) bilayers, and solar cells. While thin films show a simple response, that of bilayers is more complex. This lets us separate interface recombination from charge transfer and estimate an upper limit on the charge-carrier lifetime in the complete device. There, attempts to measure the lifetime often rely on electrical methods such as IMVS, where it can be masked by capacitive effects~\cite{Kiermasch2018RevisitingLifetimes,Kruckemeier2021ConsistentInterpretation,Ravishankar2023HowCharge}. As the optical analog of IMVS, IMPLS exhibits the same two time constants and resolves a third one. This allows distinguishing the charge-carrier lifetime from other processes in the cell.

As a case study, we use formamidinium lead triiodide (FAPbI\textsubscript{3}) solar cells with tin oxide (SnO\textsubscript{2}) as the electron transport layer (ETL) and spiro-OMeTAD (spiro) as the hole transport layer (HTL). This architecture has recently been shown to enable one of the highest power conversion efficiencies of perovskite solar cells to date~\cite{Wang2026ContinuouslyGradeddoped}. We present data with and without the addition of \textit{n}-octylammonium iodide (OAI) as a passivation agent to show the sensitivity of our methodology to reduced non-radiative recombination.

The recombination rate $R$ in a semiconductor with the charge-carrier density $n$ can be described using an empirical recombination order $\delta$ (e.g., $\delta = 1$ for first- and $\delta = 2$ for second-order processes) and $R\propto n^\delta$~\cite{Kirchartz2012MeaningReaction,Set2015AnalyticalModeling}.
Under small-signal excitation, it follows that the charge-carrier density and hence the modulated PL response $\tilde{I}_\mathrm{PL}$ can be described by a simple transfer function of the form~\cite{Set2015AnalyticalModeling}
\begin{align}
    \frac{\tilde{I}_\mathrm{PL}}{\tilde{G}}\propto \frac{1}{i\omega + \omega_\mathrm{c}}
    \label{eq:transfer_func}
\end{align}
with $\tilde{G}$ being the modulated generation rate, $\omega = 2\pi f$ the angular frequency and $\omega_\mathrm{c}$ the characteristic frequency of the system. $\omega_\mathrm{c}$ is the inverse of the time constant $\tau_\mathrm{c}$, which in this case is the small-signal or differential charge-carrier lifetime. The recombination lifetime can be obtained via~\cite{Set2015AnalyticalModeling}
\begin{align}
    \tau_\mathrm{rec}=\delta \tau_\mathrm{c}=\frac{\delta}{\omega_\mathrm{c}}
    \label{eq:rec_lifetime}
\end{align}
where $\delta$ can be extracted from $\mathrm{d} \ln \omega_\mathrm{c}/\mathrm{d} \ln  G = 1-1/\delta$ ($G$ is the steady-state generation rate). 

\begin{figure*}[ht]
    \centering

    \begin{subfigure}{0.48\textwidth}
        \centering
        \includegraphics{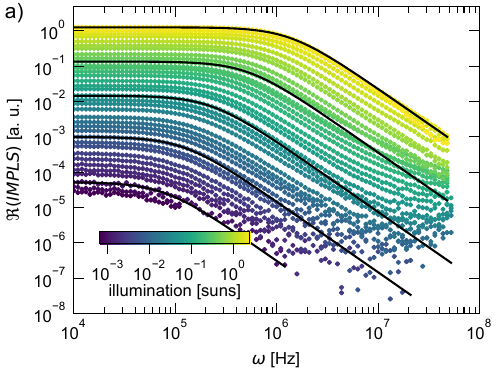}
    \end{subfigure}
    \hfill
    \begin{subfigure}{0.48\textwidth}
        \centering
        \includegraphics{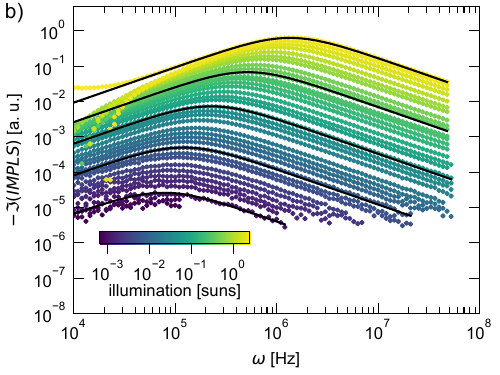}
    \end{subfigure}

    \vspace{0.5em}

    \begin{subfigure}{0.48\textwidth}
        \centering
        \includegraphics{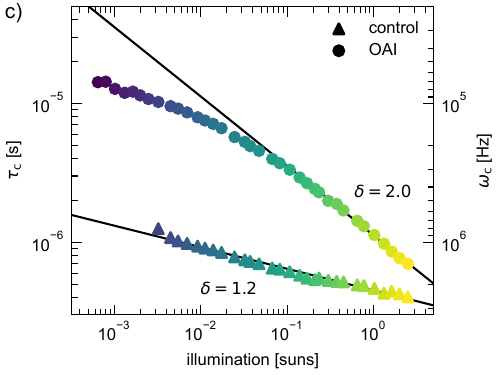}
    \end{subfigure}
    \hfill
    \begin{subfigure}{0.48\textwidth}
        \centering
        \includegraphics{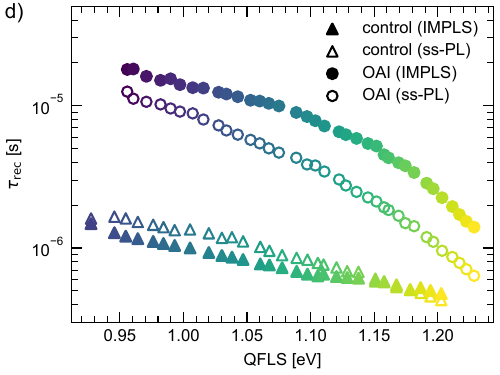}
    \end{subfigure}

    \caption{IMPLS data of perovskite thin films: Bode plots of a)~real and b)~imaginary part of the IMPLS response of an OAI-treated FAPbI\textsubscript{3} thin film for different light intensities with exemplary fit curves. c)~Extracted characteristic time constants vs light intensity with guides to the eye indicating the recombination order. d)~Recombination lifetimes from IMPLS and steady-state PL vs QFLS.}
    \label{fig:thin_film}
\end{figure*}

\Cref{fig:thin_film} a) and b) display the real and imaginary parts of the IMPLS response of an OAI-treated FAPbI\textsubscript{3} thin film, along with exemplary fit curves corresponding to \cref{eq:transfer_func}, which match the spectra over most of the frequency range. Deviations of the imaginary part at low frequencies and higher light intensities are most likely caused by the influence of mobile ions~\cite{Gillespie2025IntensityModulatedPhotoluminescence}, but can also be a result of light soaking and general measurement uncertainties at lower frequencies. The effect becomes even more pronounced when looking at the full frequency spectra (found in SI section S1, along with the Nyquist plots). Whether the deviations of the real part at lower light intensities and higher frequencies are a signature of trapped charge carriers~\cite{Moron2021AnalyticalModel,Berenguier2025ContactlessDefects} is beyond the scope of this work.

\Cref{fig:thin_film} c) shows the time constants extracted from the data via fitting of \cref{eq:transfer_func}. The OAI-treated thin film exhibits $\delta\approx2$ at higher light intensities, which reduces to about 1.2 at lower intensities. For the control film, $\delta\approx 1.2$ throughout the investigated range. The increase in recombination order upon OAI treatment is expected, as OAI is commonly used to reduce non-radiative recombination and boost the open-circuit voltage ($V_\mathrm{OC}$) of perovskite solar cells~\cite{Yuan2024ShallowDefects,Xue2026OctylammoniumIodide,Ding2024StressRegulation,Wang2019InterfacialResidual,Mahmud2021CombinedBulk}. In our case, the OAI treatment increases the $V_\mathrm{OC}$ from an average of \qty{0.98}{\volt} to \qty{1.07}{\volt} (maximum \qty{1.10}{\volt}) and the power conversion efficiency from an average of \qty{16.75}{\percent} to \qty{19.44}{\percent} (maximum \qty{21.71}{\percent}) (see SI section S3 for the $j$--$V$ curves of the investigated devices and performance statistics). The change in the recombination mechanisms of the thin film upon OAI treatment is corroborated by an increase of the photoluminescence quantum yield (PLQY) from \qty{1.8}{\percent} to \qty{7.5}{\percent} at one-sun-equivalent illumination and a change in the ideality factor $n_\mathrm{id}$: 1.5 for the control and 1.4 (1.1) at low (high) light intensities for the treated film (see SI section S2 for the determination of $n_\mathrm{id}$). \Cref{fig:thin_film} d) compares the recombination lifetimes from IMPLS (see \cref{eq:rec_lifetime}) with effective recombination lifetimes derived from the steady-state PL (see SI section S4) for different quasi-Fermi-level splitting (QFLS), determined from the relative PL intensity, a PLQY measurement at one sun, and the external quantum efficiency (EQE) of corresponding solar cells (see SI section S5). The two methods yield almost identical lifetimes for the untreated film and show a deviation of up to a factor of 2.5 for the OAI-treated film. This deviation is not surprising considering the uncertainties of the steady-state PL approach: the variation of literature values for the density of states~\cite{Staub2016BulkLifetimes,Zhou2017LowDensity}, uncertainties of the PLQY due to light soaking effects and the poorly known escape probability~\cite{Kirchartz2016ImpactPhoton,Abebe2018RigorousWaveoptical,
Kirchartz2020PhotoluminescenceBasedCharacterization,Fassl2021RevealingInternal}, and the uncertainty of the bandgap. We extracted the latter to be \qty{1.55}{\electronvolt} from the inflection point of the EQE~\cite{Kruckemeier2020HowReport} of the corresponding solar cells.

\begin{figure}[ht]
    \centering
    \includegraphics{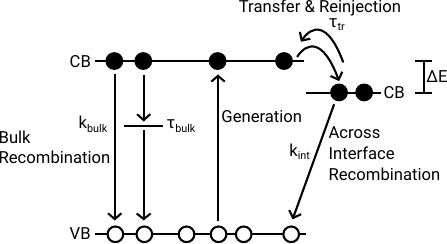}
\caption{Scheme illustrating the different processes in a perovskite/ETL bilayer.}
\label{fig:scheme-bulk-tl}
\end{figure}
\begin{figure*}[ht]
    \centering

    \begin{subfigure}{0.3\textwidth}
        \centering
        \includegraphics[width=\linewidth]{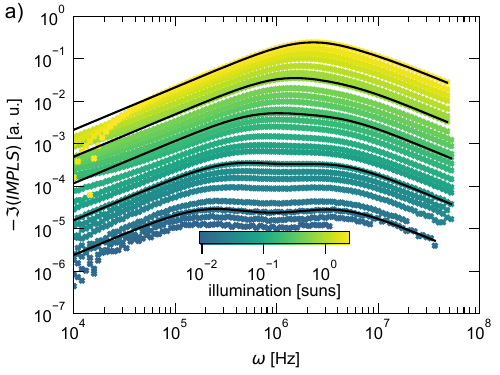}
    \end{subfigure}
    \hfill
    \begin{subfigure}{0.3\textwidth}
        \centering
        \includegraphics[width=\linewidth]{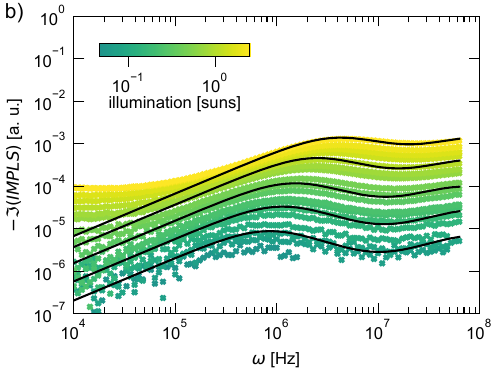}
    \end{subfigure}
    \hfill
    \begin{subfigure}{0.3\textwidth}
        \centering
        \includegraphics[width=\linewidth]{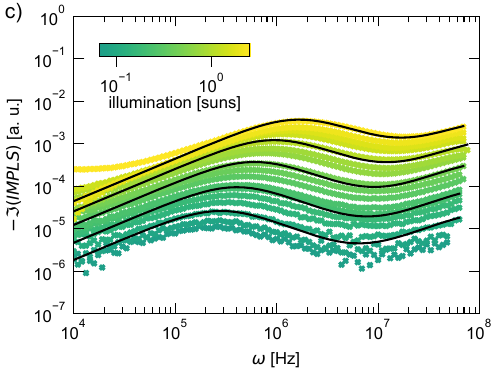}
    \end{subfigure}

    \vspace{0.5em}

    \begin{subfigure}{0.3\textwidth}
        \centering
        \includegraphics[width=\linewidth]{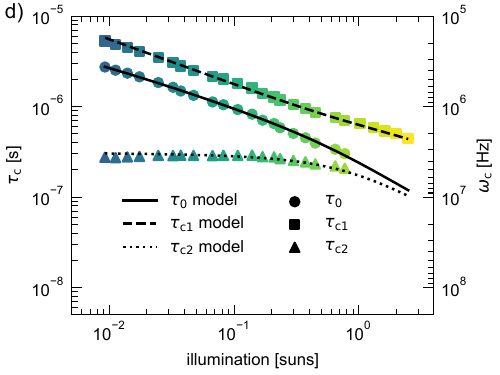}
    \end{subfigure}
    \hfill
    \begin{subfigure}{0.3\textwidth}
        \centering
        \includegraphics[width=\linewidth]{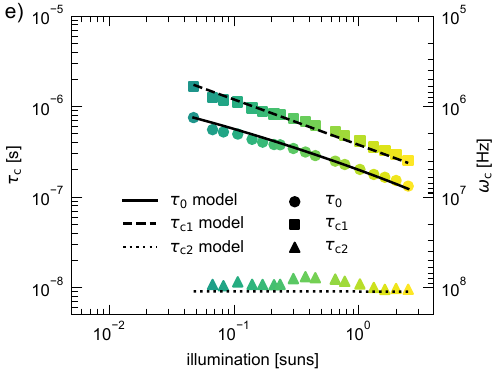}
    \end{subfigure}
    \hfill
    \begin{subfigure}{0.3\textwidth}
        \centering
        \includegraphics[width=\linewidth]{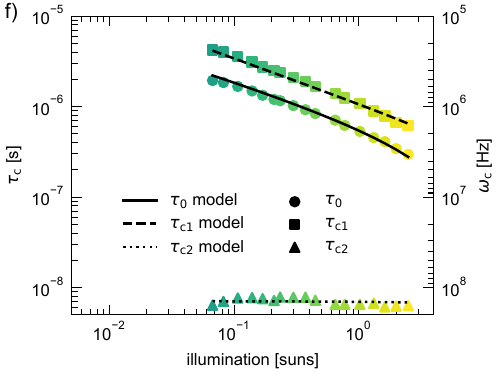}
    \end{subfigure}

    \caption{IMPLS data of perovskite/TL bilayers: Bode plots of the imaginary part of the IMPLS response of a)~SnO\textsubscript{2}/FAPbI\textsubscript{3}, b)~FAPbI\textsubscript{3}/spiro and c)~FAPbI\textsubscript{3}/OAI/spiro with exemplary fit lines of \cref{eq:transfer-func-bulk-tl}. Extracted time constants from a)-c) along with lines corresponding to a fit of the kinetic model for the case of d)~SnO\textsubscript{2}/FAPbI\textsubscript{3}, e)~FAPbI\textsubscript{3}/spiro and f)~FAPbI\textsubscript{3}/OAI/spiro.}
    \label{fig:bulk-TL}
\end{figure*}

Having established that we can reliably probe the charge-carrier lifetime in thin films, we turn our attention to perovskite/TL stacks (bilayers). In contrast to the thin films, the IMPLS response of the perovskite on SnO\textsubscript{2} and the perovskite with spiro shows two distinct time constants or characteristic frequencies, visible as two peaks in the imaginary part (see \Cref{fig:bulk-TL}). For the case of the SnO\textsubscript{2}/FAPbI\textsubscript{3} stack (a)), this is only visible at low light intensities, whereas the FAPbI\textsubscript{3}/spiro stacks without (b)) and with OAI (c)) exhibit two peaks at all intensities, with the faster one even exceeding the practical frequency limit of our setup ($\omega \approx \qty{50}{\mega\hertz}$). To explain this shape, one needs to explicitly account for the charge carriers in the TL. \Cref{fig:scheme-bulk-tl} illustrates the different processes happening in a perovskite/TL stack: generation, first- and second-order bulk recombination, transfer to the TL, reinjection from the TL to the bulk and across-interface recombination. Similarly to previous reports, we treat the system as zero-dimensional and describe it by the following set of coupled rate equations, also called the kinetic model~\cite{Hutter2017ChargeTransfer,Lee2024UnravelingLoss,Aalbers2026LowTemperatures,Zhao2024LongLivedCharge}, here shown assuming an absorber/ETL bilayer. (A corresponding set of equations can be written for the case of an absorber/HTL stack.)
\begin{align}
    \frac{dn}{dt}&=G-k_\mathrm{bulk}np-\frac{n}{\tau_\mathrm{bulk}}-\frac{1}{\tau_\mathrm{tr}}\left(n-n_\mathrm{TL}\beta\right) \label{eq:nbulk}\\
    \frac{dn_\mathrm{TL}}{dt}&=\frac{1}{\tau_\mathrm{tr}\alpha}\left(n-n_\mathrm{TL}\beta\right)- k_\mathrm{int}n_\mathrm{TL}p\label{eq:ntl}\\
    p &= \alpha n_\mathrm{TL}+n \label{eq:charge_neutrality}
\end{align} 
where $n$ and $n_\mathrm{TL}$ are the densities of electrons in the bulk and the TL, and $t$ is time. The density of holes in the absorber $p$ is simply set by charge neutrality. Besides generation $G$, bulk recombination (either first-order, described by the bulk lifetime $\tau_\mathrm{bulk}$, or second-order, with the coefficient $k_\mathrm{bulk}$) and across-interface recombination with the coefficient $k_\mathrm{int}$ may occur. The transfer of charge carriers between bulk and TL is described by the transfer time, which is the inverse of the rate at which electrons are transferred to the TL $\tau_\mathrm{tr}=1/k_\mathrm{tr}$. It depends on the interface energetics and the mobility in the absorber and TL. 

The back-transfer of electrons from the TL to the bulk is scaled by a prefactor 
\begin{align}
\beta &= \exp\left(-\frac{\Delta E}{k_\mathrm{B}T}+\frac{n_\mathrm{TL}}{n_\mathrm{\Phi}}\right) \label{eq:beta}
\end{align}
which in turn depends on two mechanisms: first, thermally activated electron transfer set by the energy offset $\Delta E$ between the conduction bands of the absorber and TL, the temperature $T$, and the Boltzmann constant $k_\mathrm{B}$; second, electrostatic effects. The importance of these for modeling bilayers is well known~\cite{Butler-Caddle2024DistinguishingCarrier,Kruckemeier2021UnderstandingTransient,Lee2024UnravelingLoss}. Instead of solving the Poisson equation in a one-dimensional drift-diffusion simulation, we use the following simplification: when $n_\mathrm{TL}\gg n$, the electrons in the TL gain electrostatic potential relative to those in the bulk, accelerating back-transfer. We write this as proportional to $\exp\left(\frac{n_\mathrm{TL}}{n_\mathrm{\Phi}}\right)$ with $n_\mathrm{\Phi}$ the TL electron density at which the TL electrons have gained $k_\mathrm{B}T$ in potential energy. The exponent in \cref{eq:beta} can therefore be understood as an effective energy offset. For simplicity, we assume that the electric field in the perovskite is largely screened by mobile ions~\cite{Ravishankar2024DiscerningRise,Ravishankar2023HowCharge,Thiesbrummel2024IoninducedField,Diethelm2025ProbingIonic} and that charge-carrier transfer to the TL and interfacial recombination are unaffected by the charge separation. The thickness ratio $\alpha = d_\mathrm{TL}/d$ (with \qtylist{600;30;300}{\nano\meter} for the perovskite, SnO\textsubscript{2} and spiro, respectively) accounts for the TL typically being much thinner than the absorber, so that the charge-carrier density in the TL rises more steeply per transferred charge than the density in the perovskite decreases.

Writing \cref{eq:nbulk,eq:ntl} as a matrix equation and linearizing around the steady state, we derive a transfer function of the shape
\begin{align}
    \frac{\tilde{I}_\mathrm{PL}}{\tilde{G}}\propto \frac{i\omega+\omega_{0}}{(i\omega+\omega_\mathrm{c1})(i\omega+\omega_\mathrm{c2})} \label{eq:transfer-func-bulk-tl}
\end{align}
The characteristic frequencies $\omega_{0}$, $\omega_\mathrm{c1}$, $\omega_\mathrm{c2}$ of the system (each of them the reciprocal of the time constants $\tau_{0}$, $\tau_\mathrm{c1}$, $\tau_\mathrm{c2}$, respectively) are rather unintuitive functions of the parameters in \cref{eq:nbulk,eq:ntl}, as derived in SI section S6. 

\Cref{fig:bulk-TL} a)--c) show the imaginary part of the IMPLS response of the three perovskite/TL stacks, with fits of \cref{eq:transfer-func-bulk-tl} at one illumination each (see SI section S7 for the real part), resulting in the corresponding time constants $\tau_i(G)$. We attribute the deviations at lower frequencies to mobile ions, as discussed for the thin films. Panels d)--f) compare the extracted $\tau_i(G)$ to a simultaneous fit of the kinetic model to these time constants and the relative steady-state PL intensity above 0.1 sun (for details on the fitting see SI section S8). The kinetic model reproduces both, and we therefore consider it an adequate description of the interplay between charge transfer and recombination. The shape of the transfer function \cref{eq:transfer-func-bulk-tl} also explains why, as in the case of the SnO\textsubscript{2}/FAPbI\textsubscript{3} bilayer (\Cref{fig:bulk-TL} a) and d)) at high light intensities, only one peak in the imaginary part might be visible: if $\omega_0$ is close to either $\omega_\mathrm{c1}$ or $\omega_\mathrm{c2}$, it cancels the respective one, and the transfer function simplifies to the shape described in \cref{eq:transfer_func}. A single peak in the imaginary part (or in equivalent time-domain methods, a monoexponential decay), therefore, does not guarantee that the result reflects the recombination lifetime.

As a zero-dimensional approximation, the kinetic model neglects the influence of spatial transport, nonuniform generation, doping in the TLs, and possible band bending at the interface. In contrast to other works~\cite{Hutter2017ChargeTransfer,Lee2024UnravelingLoss,Aalbers2026LowTemperatures,Zhao2024LongLivedCharge}, we do not explicitly account for trapped charges. Since our model already captures the observed trends, adding a fourth state to the system would likely lead to overparameterization. SI section S9 demonstrates that $k_\mathrm{int}$ and $\tau_\mathrm{tr}$ are well constrained. As the bulk recombination rates are not dominant, $\tau_\mathrm{bulk}$ and $k_\mathrm{bulk}$ remain mostly undetermined by our fits.

For the SnO\textsubscript{2}/FAPbI\textsubscript{3} as well as the FAPbI\textsubscript{3}/spiro stacks, $\omega_0=1/\tau_0$ depends strongly on light intensity. This requires treating recombination across the interface as a second-order process ($R_\mathrm{int}\propto pn_\mathrm{TL}$ for the case of an ETL), as is sometimes~\cite{Aalbers2026LowTemperatures,Lee2024UnravelingLoss}, but not always~\cite{Hutter2017ChargeTransfer,Zhao2024LongLivedCharge}, assumed. Treating it as a first-order process ($R_\mathrm{int}\propto p$) instead makes $\omega_0$ much less dependent on light intensity, and the model fails to reproduce the data (see SI section S10). Hence, the surface recombination cannot be described by a single surface recombination velocity $S$ -- as is commonly done -- because $S$ becomes a function of charge-carrier density. The definition of $S$ also varies in the literature and studies do not always discriminate between charge extraction and recombination~\cite{Aalbers2026LowTemperatures,Kruckemeier2021UnderstandingTransient,Gillespie2024SiliconInspiredAnalysis,Sproul1994DimensionlessSolution,Wang2019ReducingSurface}. While the fit of the kinetic model (see \Cref{tab:parameters}) constrains $k_\mathrm{int}$ and $\tau_\mathrm{tr}$ well, a more accurate determination of $\Delta E$ would require temperature-dependent measurements. Without these, $\Delta E$ can be significantly increased without the fit getting much worse (see SI section S9), leading to uncertainties in calculating the interface recombination velocity. Interface recombination and charge extraction may also depend on the measurement condition: in a transient experiment, the TL can be considered free of excess carriers before the laser pulse, whereas in a steady-state scenario, the TL may be significantly charged. All of this complicates the comparison with reported values, for which we provide a short overview in SI section S11. 

\begin{table}[ht]
\small
\centering
\caption{Interfacial parameters of the different perovskite/TL stacks obtained from a fit of the kinetic model to the IMPLS data.}
\label{tab:parameters}
\begin{tabular}{lccc}
    \hline
    Sample & SnO\textsubscript{2}/ & FAPbI\textsubscript{3}/ & FAPbI\textsubscript{3}/ \\
    &FAPbI\textsubscript{3}&spiro&OAI/spiro\\
    \hline
    $k_\mathrm{int}\,[\unit{\centi\meter\cubed\per\second}]$ & $\num{3e-10}$ & \num{6e-10} & \num{8e-11}\\
    $\tau_\mathrm{tr}$\,[\unit{\nano\second}] & $\num{310} $ & \num{10} & \num{8}\\
    $\Delta E$ \,[\unit{\electronvolt}]&0.19&0.18&0.24\\
    \hline
\end{tabular}
\end{table}

 The SnO\textsubscript{2}/FAPbI\textsubscript{3} bilayer exhibits comparatively slow dynamics. We extract a charge-transfer time of \qty{310}{\nano\second} compared with only \qty{10}{\nano\second} for the FAPbI\textsubscript{3}/spiro bilayers. The latter value corresponds to a transfer rate of $k_\mathrm{tr}=\qty{1e8}{\per\second}$ and lies in the range of previously reported values of $k_\mathrm{tr}=\qtyrange{5e6}{8e8}{\per\second}$ obtained from time-resolved microwave conductivity and time-resolved surface photovoltage on various perovskite/spiro bilayers~\cite{Hutter2017ChargeTransfer,Caselli2022QuantifyingCharge,Iqbal2023InterfaceModification}. $\tau_\mathrm{tr}\approx\qty{10}{\nano\second}$ is very close to the timescale for diffusion to the interface $\tau_\mathrm{diff}=4/D\cdot (d/\pi)^2$~\cite{Sproul1994DimensionlessSolution}. With the ambipolar diffusion constant $D$ assumed to be \qty{2.6e-1}{\centi\meter\squared\per\second} -- corresponding to a mobility of \qty{10}{\centi\meter\squared\per\volt\per\second} -- this results in $\tau_\mathrm{diff}=\qty{6}{\nano\second}$. The charge transfer to the spiro can therefore be considered efficient and is limited primarily by the hole diffusivity in the perovskite. Upon OAI treatment, the interfacial recombination coefficient of the spiro interface is reduced by almost an order of magnitude; however, the fit unexpectedly yields a larger energy offset, likely reflecting that $\Delta E$ is not well constrained. This increased apparent energy offset leads to an increase in the density of holes in the TL, which partly counteracts the reduced $k_\mathrm{int}$. As a result, the steady-state PL intensity of the OAI-treated stack during the IMPLS experiment is about a factor of two larger than that of the control.

For the SnO\textsubscript{2}/FAPbI\textsubscript{3} bilayer at light intensities above 1 sun, the electrostatic term in \cref{eq:beta} becomes important (see SI section S12 for a fit without the term): the negative charge accumulated in the SnO\textsubscript{2} leads to increased electron reinjection into the bulk, so that forward transfer and back-injection approach balance. While the bulk is essentially starved of electrons at low injection, the electron density now approaches the hole density. This effect resembles the reduced charge extraction observed when a bias light is applied in addition to the pulsed excitation~\cite{Zhao2024LongLivedCharge} and has also been referred to as the Coulomb bottleneck~\cite{Butler-Caddle2024DistinguishingCarrier}. It is not visible in the bilayers with spiro, probably because it is much thicker; it can accommodate many more holes before a significant space charge builds up. 

A TL with a significant energy offset and thickness will lead to a high asymmetry in electron and hole densities in the absorber. This asymmetry is (at least partially) removed when a second TL counteracts the first one. This, and the influence of the contacts, complicates the transfer of the extracted values from the bilayers to the full device. We therefore extend \crefrange{eq:nbulk}{eq:charge_neutrality} by a fourth equation for the charge-carrier density in the second TL, so that the equations describe an ETL/perovskite/HTL stack (see SI section S13). Solving the system numerically with the fit values from the bilayers, assuming a generation rate equivalent to one sun and neglecting bulk recombination, we arrive at upper limits for interface-dominated recombination lifetimes: \qty{200}{\nano\second} (control) and \qty{750}{\nano\second} (OAI-treated).


On full solar cells, IMPLS becomes the optical analog of IMVS. The former directly measures the response of the charge-carrier density and QFLS, or the implied voltage $V_\mathrm{int}$, upon modulation of the excitation, whereas the latter measures the response of the external voltage at the electrodes, $V_\mathrm{elec}$. To relate the two techniques, we use the framework by Ravishankar et al.~\cite{Ravishankar2023HowCharge,Ravishankar2024DiscerningRise}, who developed an analytical model describing IMVS on perovskite solar cells. The main assumptions of the equivalent circuit model are an intrinsic field-free absorber (screened by mobile ions) with a uniform charge-carrier distribution; a constant electric field across the TLs; and no significant band offset between the absorber conduction and valence band and the respective bands of the TLs. The last two are not strictly valid here, as SnO\textsubscript{2} and spiro are generally doped, and the bilayer measurements indicate non-zero energy offsets. However, we will show that the model still describes our data accurately.

Under these premises, the IMVS transfer function takes the general shape (see~\cite{Ravishankar2023HowCharge,Ravishankar2024DiscerningRise} and SI section S14)
\begin{align}
    \frac{\tilde{V}_\mathrm{OC}}{\tilde{G}}\propto \frac{1}{(i\omega+\omega_\mathrm{c1})(i\omega+\omega_\mathrm{c2})} \label{eq:imvs_transfer}
\end{align}
The characteristic frequencies depend on the elements of the equivalent circuit: the recombination resistance $R_\mathrm{rec}$ quantifies the loss of carriers due to recombination; the chemical capacitance $C_\mu$ is a measure of the charge-carrier density; $C_\mathrm{g}$ is the geometric capacitance and $R_\mathrm{exc}$ describes the voltage drop across the TLs. Related to these circuit elements are the effective charge-carrier lifetime $\tau_\mathrm{eff}$, the charge-carrier exchange time $\tau_\mathrm{exc}$ and the electrode charging time $\tau_\mathrm{elec}$. When solving the same system for the observable $V_\mathrm{int}$ rather than $V_\mathrm{elec}$, we obtain a transfer function with the same general shape as \cref{eq:transfer-func-bulk-tl}, which is of great benefit: IMVS maps the three unknowns $\tau_\mathrm{eff}$, $\tau_\mathrm{exc}$, $\tau_\mathrm{elec}$ onto two characteristic frequencies, so the unknowns can only be derived with additional measurements or specific assumptions, while IMPLS maps three unknowns onto three frequencies. This opens up the possibility of disentangling the effective charge-carrier lifetime in complete solar cells from charge-carrier exchange and capacitive effects, which is essential for a correct determination of this important device parameter~\cite{Kiermasch2018RevisitingLifetimes,Kruckemeier2021ConsistentInterpretation,Krogmeier2018QuantitativeAnalysis}. The three time constants are defined by the circuit elements and can be obtained from the characteristic frequencies (see~\cite{Ravishankar2023HowCharge,Ravishankar2024DiscerningRise} and SI section S14):
\begin{align}
    \tau_\mathrm{eff} &= R_\mathrm{rec}C_\mu
                    &&= \frac{\omega_0}{\omega_\mathrm{c1}\omega_\mathrm{c2}}
                        \label{eq:tau_rec}\\
    \tau_\mathrm{exc} &= R_\mathrm{exc}C_\mu 
                    &&= \frac{1}{\omega_\mathrm{c1}+\omega_\mathrm{c2}-\omega_0-\dfrac{\omega_\mathrm{c1}\omega_\mathrm{c2}}{\omega_0}}
                    \label{eq:tau_exc}\\
    \tau_\mathrm{elec} &= R_\mathrm{exc}C_\mathrm{g}
                    &&= \frac{1}{\omega_0}
                        \label{eq:tau_elec}
\end{align}
$\tau_\mathrm{eff}$ is a small-signal lifetime that needs to be corrected via \cref{eq:rec_lifetime}. $\tau_\mathrm{exc}$ is a full device-level parameter that describes how long it takes a charge carrier to travel from the absorber to the contact, whereas the previously discussed $\tau_\mathrm{tr}$ only quantifies the transfer between the absorber and the TL. 

\begin{figure*}[ht]
    \centering
    \begin{subfigure}{0.48\textwidth}
        \centering
        \includegraphics[width=\linewidth]{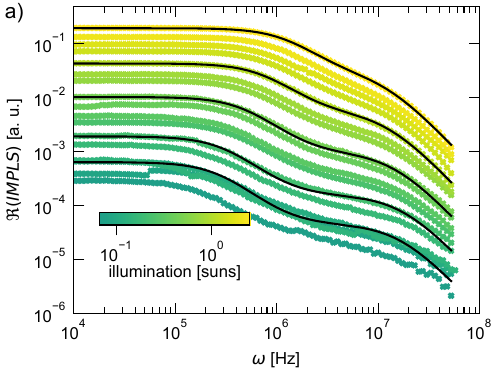}
    \end{subfigure}
    \hfill
    \begin{subfigure}{0.48\textwidth}
        \centering
        \includegraphics[width=\linewidth]{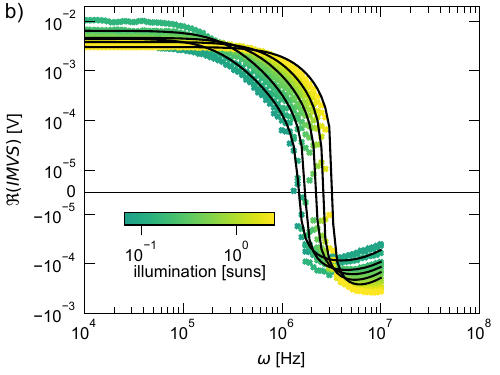}
    \end{subfigure}

    \vspace{0.5em}

    \begin{subfigure}{0.48\textwidth}
        \centering
        \includegraphics[width=\linewidth]{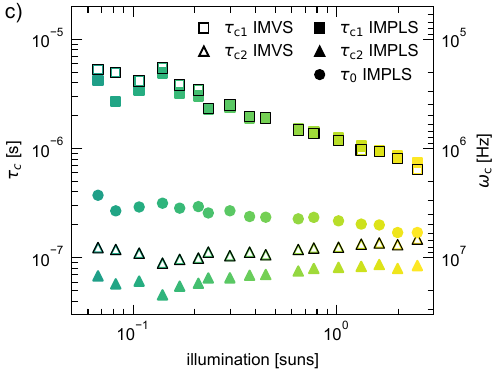}
    \end{subfigure}
    \hfill
    \begin{subfigure}{0.48\textwidth}
        \centering
        \includegraphics[width=\linewidth]{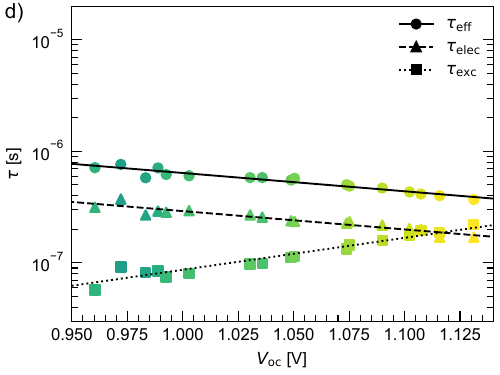}
    \end{subfigure}

    \caption{IMPLS and IMVS data of the SnO\textsubscript{2}/FAPbI\textsubscript{3}/OAI/spiro solar cell: real part of the a) IMPLS and b) IMVS response, c) extracted characteristic frequencies from IMVS and IMPLS for different light intensities, d) recombination lifetimes, exchange times and electrode charging times calculated from the characteristic frequencies of the IMPLS vs $V_\mathrm{OC}$ with fit lines as guide to the eye.}
    \label{fig:cells}
\end{figure*}

In the following, we discuss the IMPLS and IMVS results of complete solar cells, focusing on the OAI-treated device (see SI section S15 for the control device). \Cref{fig:cells} a) and b) show IMPLS and IMVS measurements on the treated cell with fits to \cref{eq:transfer-func-bulk-tl} and \cref{eq:imvs_transfer}, respectively. To highlight the different shape of the transfer functions, the real part is displayed (see SI section S16 for the imaginary part). In both cases the model reproduces the spectra over the relevant frequency range. The sign change of the real part in IMVS is an important signature of \cref{eq:imvs_transfer} and demonstrates the information loss of the common approach of extracting the characteristic frequency by simply taking the peak position of the imaginary part or the maximum of the Nyquist plot. This sign change is suppressed by $\omega_0$ in IMPLS, where the real part stays positive and a second shoulder appears.

\Cref{fig:cells} c) displays the extracted time constants from IMVS and IMPLS for different light intensities. The slow time constant $\tau_\mathrm{c1}$ agrees between the two techniques, whereas the fast time constant $\tau_\mathrm{c2}$ deviates by up to a factor of 2. We attribute this to limitations in the IMVS setup: long cables ($\approx \qty{2.5}{\meter}$) introduce a signal delay and stray capacitance, and the amplitude decreases steeply once $\omega \gg \omega_\mathrm{c1}$, limiting $\omega$ to a usable maximum of \qty{10}{\mega\hertz}.

Using \crefrange{eq:tau_rec}{eq:tau_elec}, we calculate the physical parameters governing the IMPLS and IMVS response of the solar cell, shown in \Cref{fig:cells} d). The effective recombination lifetime decreases from about $\qty{760}{\nano\second}$ at 0.06 suns to about $\qty{370}{\nano\second}$ at two suns with \qty{470}{\nano\second} at one sun. This highlights the importance of careful interpretation of IMVS data -- simply taking the peak of the imaginary part results in apparent lifetimes of about \qty{1}{\micro\second}. The weak intensity dependence corresponds to $\delta=1.2$, consistent with the high ideality factor (1.7 from suns-$V_\mathrm{OC}$, 1.9 from suns-PL, see SI section S17). Assuming a negligible intrinsic carrier density, this indicates that the recombination proceeds mostly via deep traps. This matches the lower slope of $\tau_\mathrm{rec}$ at lower QFLS in \Cref{fig:thin_film} d). However, the results for the bilayers indicate that recombination scales more strongly with light intensity, which we cannot yet reconcile. When corrected for the recombination order, the OAI-treated cell shows $\tau_\mathrm{rec}\approx\qty{560}{\nano\second}$ at one sun -- below the predicted limit of \qty{750}{\nano\second}. For the control device, $\tau_\mathrm{eff}\approx \qty{190}{\nano\second}$ -- almost identical to the predicted limit --
at all investigated light intensities (corresponding to $\delta =1$, which means the differential lifetime equals the recombination lifetime). The observed increase in $\tau_\mathrm{rec}$ by a factor of 2.9 roughly matches the increased steady-state PL intensity during the IMPLS measurement at one sun by a factor of about 6.5; since PL scales with $n^2$, this would correspond to an increase in $\tau_\mathrm{rec}$ by a factor of 2.6. 

However, this increase in charge-carrier lifetime can explain neither the average $V_\mathrm{OC}$ increase of \qty{90}{\milli\volt} (see SI section S3) nor the substantial increase in fill factor (\qty{71}{\percent} to \qty{75}{\percent}). This points towards the OAI having an additional effect, such as improving energy-level alignment or charge extraction. As interface recombination leads to a greater decrease in $V_\mathrm{OC}$ than in QFLS in the bulk~\cite{Caprioglio2019RelationOpenCircuit}, and IMPLS is mostly sensitive to the modulation of the QFLS in the bulk, optical measurements may not fully capture a reduction in interface recombination. Strong surface recombination would also explain the gap between the ideality factors from suns-$V_\mathrm{OC}$ and suns-PL (1.5 vs 2.0 for the control cell, which is reduced to 1.7 vs 1.9 for the OAI-treated cell, see SI section S17 for the corresponding plots)~\cite{Caprioglio2019RelationOpenCircuit}. This would also imply that the charge-carrier density is not fully uniform, as assumed in the model. Insights into charge extraction are discussed in SI section S18.

We have investigated the charge-carrier lifetime under operating conditions in halide perovskite thin films, layer stacks and solar cells using IMPLS. On thin films, lifetimes from IMPLS deviate from steady-state PL values by up to a factor of 2.5. By deriving a transfer function for IMPLS measurements of bilayers, we demonstrate the need for caution when evaluating the data: Even when only a single peak in the imaginary part is visible, the time constant associated with that peak might not be representative of the charge-carrier lifetime. Measurements on bilayers suggested upper lifetime limits for the solar cells: \qty{200}{\nano\second} for the control and \qty{750}{\nano\second} for the treated device. We emphasize the ability of IMPLS to measure the charge-carrier lifetime in full solar cells, a clear advantage over IMVS. We determined \qty{190}{\nano\second} for the control device (just below the predicted limit) and \qty{560}{\nano\second} for the OAI-treated solar cell at one-sun-equivalent illumination, compared to \qty{500}{\nano\second} (control) and \qty{2}{\micro\second} (OAI-treated) of the
corresponding thin films at the same injection level. As demonstrated here, IMPLS can be a helpful tool throughout the processing chain for perovskite solar cells and aid in understanding the interplay between interfaces and bulk recombination. It also opens a new avenue for studying the recombination mechanisms in halide perovskites, which are still not fully understood~\cite{Stranks2026JourneyGlobal}.

\section*{Supporting Information}
Additional IMPLS and IMVS data, suns-PL and suns-$V_\mathrm{OC}$ data, $j$--$V$ curves and device statistics, EQE, mathematical derivations, fitting procedure and uncertainties, experimental methods, materials and device fabrication (PDF).

\section*{Data Availability}
The fitting code for the kinetic model, together with sample data, is available at
\url{https://github.com/deibellab/impls-bilayer-fit}.

\section*{Author Information}
\textbf{Corresponding Author:}

Carsten Deibel -- Institut für Physik, Technische Universität Chemnitz, 09126 Chemnitz, Germany; \url{orcid.org/0000-0002-3061-7234}; Email: deibel@physik.tu-chemnitz.de

\textbf{Authors:}

Constantin Bach -- Institut für Physik, Technische Universität Chemnitz, 09126 Chemnitz, Germany;

Dong Won Kim -- Leibniz-Institut für Festkörper- und Werkstoffforschung Dresden (IFW), 01069 Dresden, Germany; Chair for Emerging Electronic Technologies, Technische Universität Dresden, 01187 Dresden, Germany; \url{orcid.org/0000-0001-9945-734X}

Yitian Du -- Leibniz-Institut für Festkörper- und Werkstoffforschung Dresden (IFW), 01069 Dresden, Germany; Chair for Emerging Electronic Technologies, Technische Universität Dresden, 01187 Dresden, Germany;

Yana Vaynzof -- Leibniz-Institut für Festkörper- und Werkstoffforschung Dresden (IFW), 01069 Dresden, Germany; Chair for Emerging Electronic Technologies, Technische Universität Dresden, 01187 Dresden, Germany; \url{orcid.org/0000-0002-0783-0707}

\textbf{Notes:}
The authors declare no competing financial interest.

\section*{Acknowledgements}
The authors thank the German Research Foundation (Deutsche Forschungsgemeinschaft, DFG) for generous support within the framework of SPP 2196 project (PERFECT PVs, project no. 424216076). Funding by the DFG via the ``Responsible Electronics in the Climate Change Era -- REC2'' Cluster of Excellence (EXC 3035, Project-ID 533607596) is gratefully acknowledged. The authors also acknowledge support from the Leibniz Programme for Women Professors (Project SUPERSOL). We thank Thomas Kirchartz and Sandheep Ravishankar for fruitful discussions. Claude (Anthropic) was used as a support tool when writing the python code for evaluating the data, language editing and checking mathematical derivations for errors.

\printbibliography
\end{refsection}


\clearpage
\onecolumn
\begin{refsection} 

\renewcommand{\thesection}{S\arabic{section}}
\renewcommand{\thesubsection}{S\arabic{section}.\arabic{subsection}}
\renewcommand{\thefigure}{S\arabic{figure}}
\renewcommand{\thetable}{S\arabic{table}}
\renewcommand{\theequation}{S\arabic{equation}}
\renewcommand{\thepage}{S\arabic{page}}

\setcounter{secnumdepth}{1}
\setcounter{secnumdepth}{2}

\setcounter{figure}{0}
\setcounter{table}{0}
\setcounter{equation}{0}
\setcounter{section}{0}
\setcounter{page}{1}

\begin{center}
  {\Large\bfseries Supporting Information for:\par}
  \vspace{0.4cm}
  {\Large Operando Charge Carrier Dynamics by Intensity-Modulated Photoluminescence: From Perovskite Thin Films to Solar Cells\par}
  \vspace{0.4cm}
  {\large Constantin Bach\textsuperscript{1}, Dong Won Kim\textsuperscript{2,3}, Yitian Du\textsuperscript{2,3}, Yana Vaynzof\textsuperscript{2,3}, Carsten Deibel\textsuperscript{1}\par}
\vspace{0.4cm}
\textsuperscript{1}Institut für Physik, Technische Universität Chemnitz, 09126 Chemnitz, Germany

\textsuperscript{2}Leibniz-Institut für Festkörper- und Werkstoffforschung Dresden (IFW), 01069 Dresden, Germany

\textsuperscript{3}Chair for Emerging Electronic Technologies, Technische Universität Dresden, 01187 Dresden, Germany
\end{center}

\tableofcontents
\clearpage

    \section{IMPLS Data of Perovskite Thin Films}
    \label{sec:impls_film}
    \begin{figure}[H]
        \centering
        \begin{subfigure}{0.48\textwidth}
            \centering
            \includegraphics{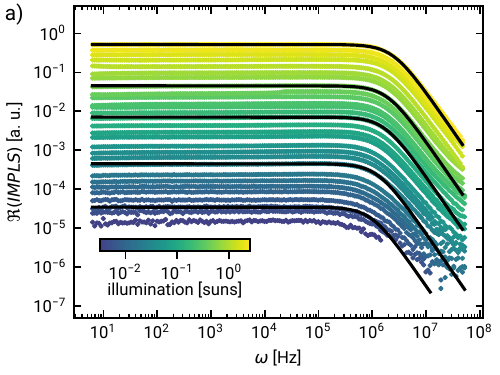}
        \end{subfigure}
        \hfill
        \begin{subfigure}{0.48\textwidth}
            \centering
            \includegraphics{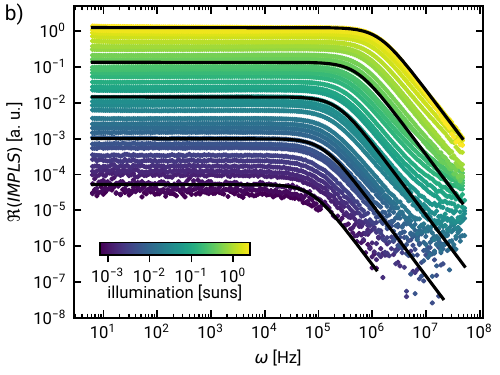}
        \end{subfigure}
        \caption{Bode plots of the real part of the IMPLS response of two FAPbI\textsubscript{3} thin films: a)~control and b)~OAI-modified with exemplary fit lines of the simple transfer function as stated in the main text.}
        \label{fig:real}
    \end{figure}

    \begin{figure}[H]
        \centering
        \begin{subfigure}{0.48\textwidth}
            \centering
            \includegraphics{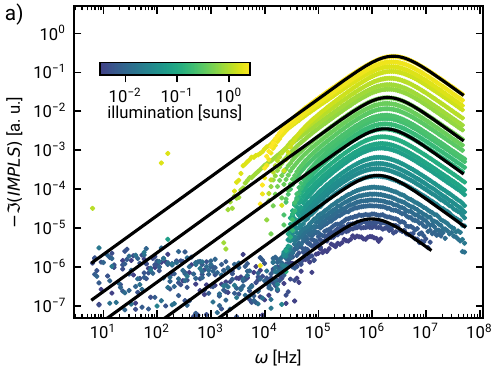}
        \end{subfigure}
        \hfill
        \begin{subfigure}{0.48\textwidth}
            \centering
            \includegraphics{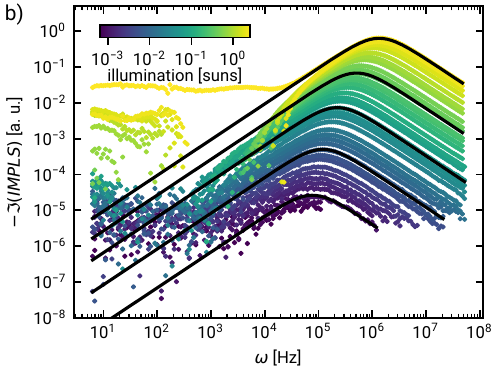}
        \end{subfigure}
        \caption{Bode plots of the imaginary part of the IMPLS response of the two measured FAPbI\textsubscript{3} thin films: a)~control, b)~OAI-modified with exemplary fit lines of the simple transfer function as stated in the main text.}
        \label{fig:imag}
    \end{figure}

    \begin{figure}[H]
        \centering
        \begin{subfigure}{0.48\textwidth}
            \centering
            \includegraphics{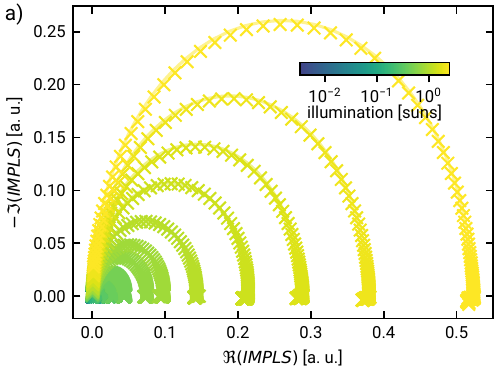}
        \end{subfigure}
        \hfill
        \begin{subfigure}{0.48\textwidth}
            \centering
            \includegraphics{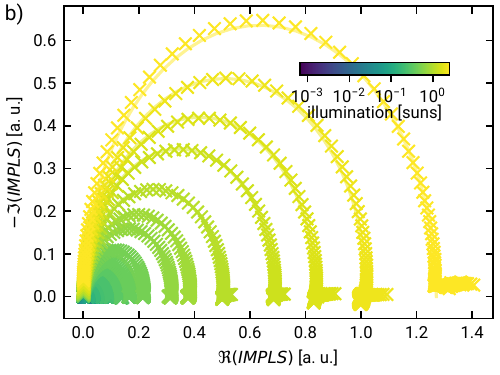}
        \end{subfigure}
        \caption{Nyquist plots of the IMPLS response of the two measured FAPbI\textsubscript{3} thin films: a)~control, b)~OAI-modified with fit lines of the simple transfer function as stated in the main text. The OAI-treated film shows some light soaking at high light intensities, which leads to a decrease in the real part of the IMPLS signal at low frequencies.}
        \label{fig:nyquist}
    \end{figure}

    \section{Ideality Factor from Photoluminescence Measurements}
    \label{sec:optical_ideality}

\begin{figure}[H]
        \centering
        \begin{subfigure}{0.48\textwidth}
            \centering
            \includegraphics{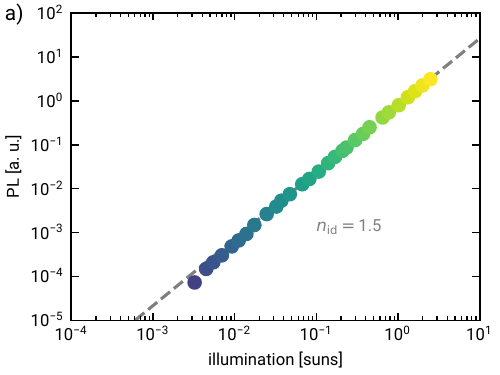}
        \end{subfigure}
        \hfill
        \begin{subfigure}{0.48\textwidth}
            \centering
            \includegraphics{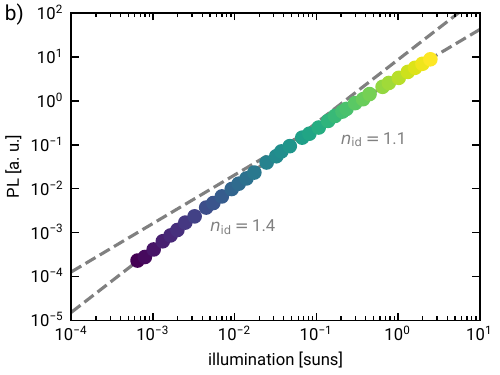}
        \end{subfigure}
        \caption{Suns-PL plots of the two measured FAPbI\textsubscript{3} thin films: a)~control and b)~OAI-modified with linear fit lines determining the ideality factor.}
        \label{fig:suns_pl}
    \end{figure}

    To determine the ideality factor from PL, we determine the relative steady-state PL intensity $I_\mathrm{PL}$ with varying illumination intensity/generation rate $G$ (see \cref{fig:suns_pl}). The ideality factor is calculated by~\cite{Sarritzu2017OpticalDetermination,Calado2019IdentifyingDominant,Caprioglio2019RelationOpenCircuit}
    \begin{align}
        n_\mathrm{id,PL} = \frac{d \ln\left(I_\mathrm{PL}\right)}{d \ln\left(G\right)}
    \end{align}

    \section{J-V Curves and Device Performance}
    \label{sec:device_performance}
    \begin{figure}[H]
        \centering

        \begin{subfigure}{0.48\textwidth}
            \centering
            \includegraphics{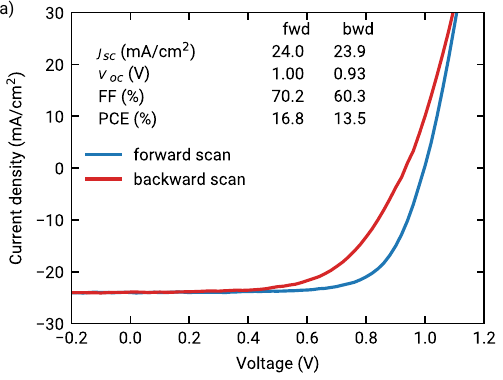}
        \end{subfigure}
        \hfill
        \begin{subfigure}{0.48\textwidth}
            \centering
            \includegraphics{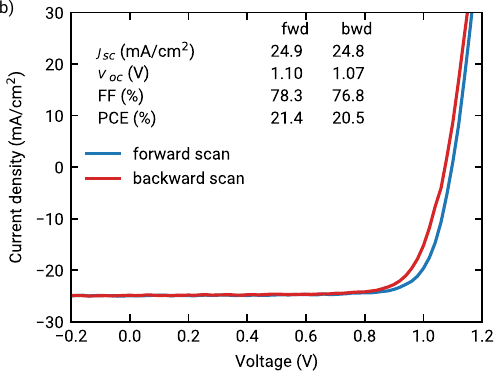}
        \end{subfigure}

        \caption{$j$--$V$ curves of the devices used for IMVS and IMPLS measurements: a) control and b) OAI-treated.}
        \label{fig:j-V}
    \end{figure}

    \begin{figure}[H]
        \centering

        \begin{subfigure}{0.48\textwidth}
            \centering
            \includegraphics{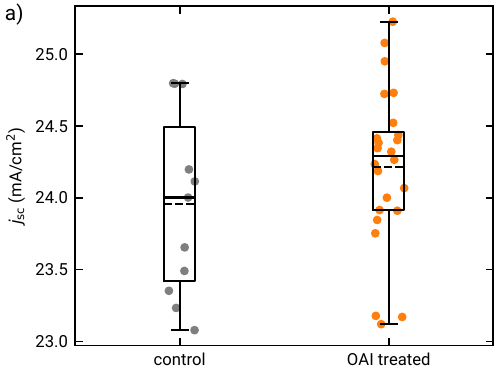}
        \end{subfigure}
        \hfill
        \begin{subfigure}{0.48\textwidth}
            \centering
            \includegraphics{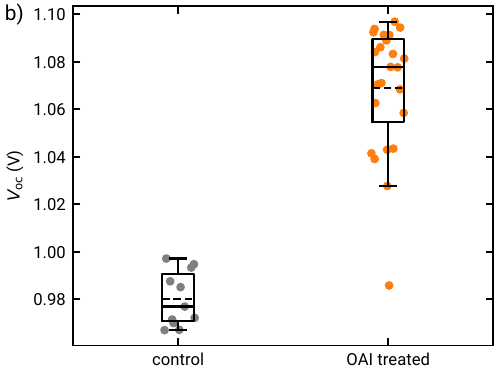}
        \end{subfigure}

        \vspace{0.5em}

        \begin{subfigure}{0.48\textwidth}
            \centering
            \includegraphics{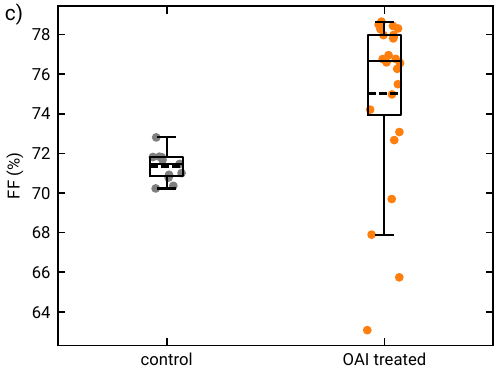}
        \end{subfigure}
        \hfill
        \begin{subfigure}{0.48\textwidth}
            \centering
            \includegraphics{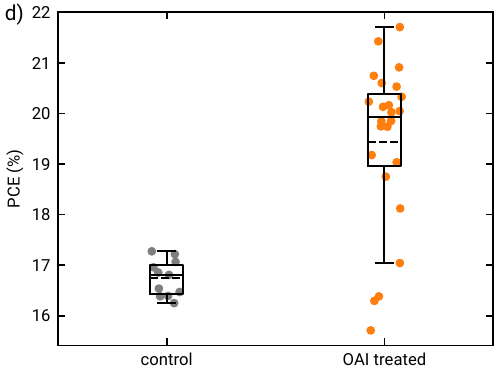}
        \end{subfigure}

        \caption{Box-plots of the device performance metrics: a) short circuit current density, b) open-circuit voltage, c) fill factor, d) power conversion efficiency. Sample size is 11 for the control and 24 for the OAI-treated devices. Box shows interquartile range, solid (dashed) line inside the box the median (mean), whiskers extend to highest/lowest data point inside 1.5 times the interquartile range. All data points are taken from the backward scan.}
        \label{fig:device-metrics}
    \end{figure}

    \section{Effective Recombination Lifetimes from Steady-State Photoluminescence}
    \label{sec:ss_lifetime}

    The Quasi-Fermi-Level-Splitting $\mathrm{QFLS}$ is related to the photoluminescence quantum yield $\mathrm{PLQY}$ via~\cite{Sarritzu2017OpticalDetermination}
    \begin{align}
        \mathrm{QFLS} = k_\mathrm{B}T\cdot \ln\left(\mathrm{PLQY}\cdot \frac{J_\mathrm{gen}}{J_{0,\mathrm{rad}}}\right),
    \end{align}
    with $k_\mathrm{B}$ Boltzmann constant, $T$ temperature and $J_\mathrm{gen}$ generation current density. According to Rau's reciprocity relation the dark saturation current density can be calculated by~\cite{Rau2007ReciprocityRelation}
    \begin{align}
        J_{0,\mathrm{rad}} = q\int_{-\infty}^{\infty}~\mathrm{EQE}(E)~\Phi_\mathrm{BB}(E)~dE,
    \end{align}
    where $q$ is the elementary charge, $\mathrm{EQE}(E)$ is the $\mathrm{EQE}$ of a corresponding solar cell and $\Phi_\mathrm{BB}$ the black body spectrum. Assuming a symmetric density of electrons and holes $n$ and the intrinsic charge carrier density $n_\mathrm{i} = \sqrt{N_\mathrm{C} N_\mathrm{V}} \cdot \exp\left(\frac{-E_\mathrm{g}}{2k_\mathrm{B}T}\right)$ ($N_\mathrm{C}$, $N_\mathrm{V}$ density of states in the conduction and valence band, respectively, both assumed to be $\qty{2.21E18}{\per\centi\meter\cubed}$~\cite{Staub2016BulkLifetimes}, bandgap $E_\mathrm{g}=\qty{1.55}{\electronvolt}$), the mass action law reads
    \begin{align}
        \mathrm{QFLS}=k_\mathrm{B}T\ln\left(\frac{n^2}{n_\mathrm{i}^2}\right).
    \end{align}
    With the generation rate $G$ an effective charge carrier lifetime
    \begin{align}
        \tau_\mathrm{rec}=\frac{n}{G} \label{eq:simple_tau}
    \end{align}
    is determined.

    PLQY is measured under one sun equivalent photon flux and all other data points are calculated by the relative change in steady-state excitation intensity and steady-state PL intensity during the IMPLS measurements.

    \section{EQE Measurements of the Full Solar Cells}
    \label{sec:eqe}
    \begin{figure}[H]
        \centering
        \includegraphics{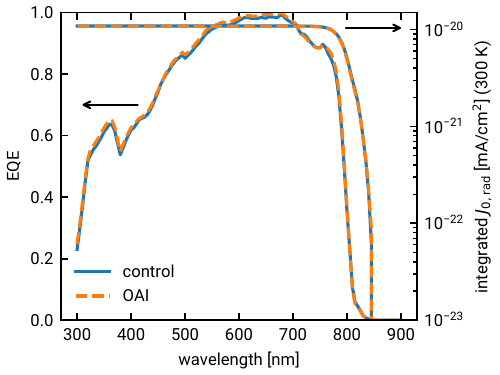}
        \caption{External quantum efficiency and integrated radiative dark saturation current density of an OAI-treated and control solar cell.}
        \label{fig:eqe}
    \end{figure}

    \section{IMPLS Transfer Function for a Perovskite/Transport Layer Stack}
    \label{sec:derivation_bulk_tl}
    Rate equations as in the main text:
    \begin{align}
        \frac{dn}{dt}&=G-k_\mathrm{bulk}np-\frac{n}{\tau_\mathrm{bulk}}-\frac{1}{\tau_\mathrm{tr}}\left(n-n_\mathrm{TL}\beta\right)\label{eq:dndt}\\
        \frac{dn_\mathrm{TL}}{dt}&=\frac{1}{\tau_\mathrm{tr}\alpha}\left(n-n_\mathrm{TL}\beta\right)- k_\mathrm{int}n_\mathrm{TL}p \label{eq:dntldt}\\
        p &= \alpha n_\mathrm{TL}+n \label{eq:SI_charge_neutrality}\\
        \beta &= \exp\left(-\frac{\Delta E}{k_\mathrm{B}T}+\frac{n_\mathrm{TL}}{n_\Phi}\right) \label{eq:SI_beta}
    \end{align}
    Defining
    \begin{align}
        \beta_\mathrm{DC}=\exp\left(-\frac{\Delta E}{k_\mathrm{B}T}+\frac{n_\mathrm{TL,DC}}{n_\Phi}\right)
    \end{align}

    Linearizing with (note that here $G_\mathrm{DC}$, $n_\mathrm{DC}$, $n_\mathrm{TL,DC}$ denote the steady-state generation rate, charge carrier density in the bulk and in the TL and $\delta$ denotes a small perturbation, not the recombination order)
    \begin{align}
        n &= \left(n_\mathrm{DC}+\delta n\right)\\
        n_\mathrm{TL}&=\left(n_\mathrm{TL,DC}+\delta n_\mathrm{TL}\right)\\
        G&=\left(G_\mathrm{DC}+\delta G\right)
    \end{align}
    neglecting all higher order terms, writing the system as a matrix equation
        \begin{align}
        \frac{d}{dt}\begin{bmatrix}
        \delta n\\ \delta n_\mathrm{TL}
        \end{bmatrix}=\mathbf{A}\begin{bmatrix}
        \delta n\\\delta n_\mathrm{TL}
        \end{bmatrix}+\begin{bmatrix}
        \delta G\\0
        \end{bmatrix}
    \end{align}
    and using the ansatz
    \begin{align}
        \delta n(t)&=\tilde{n}e^{i\omega t}\\
        \delta n_\mathrm{TL}(t)&=\tilde{n}_\mathrm{TL}e^{i\omega t}\\
        \delta G &=\tilde{G}e^{i\omega t}
    \end{align}
    the derivative becomes
    \begin{align}
        \frac{d}{dt} = i \omega
    \end{align}
    and therefore the matrix equation reads
    \begin{align}
        i\omega\begin{bmatrix}\tilde{n}\\\tilde{n}_\mathrm{TL}
        \end{bmatrix}&=\mathbf{A}\begin{bmatrix}
        \tilde{n}\\\tilde{n}_\mathrm{TL}
        \end{bmatrix}+\begin{bmatrix}
        \tilde{G}\\0
        \end{bmatrix}
    \end{align}
    With
    \begin{align}
        \mathbf{M}&=i\omega \begin{bmatrix}
            1 & 0 \\ 0 & 1
        \end{bmatrix} - \textbf{A} \\
        &=\begin{bmatrix}
        i\omega+a&b\\c &i\omega+d
        \end{bmatrix}\\
        a &=\left(k_\mathrm{bulk}2n_\mathrm{DC}+k_\mathrm{bulk}\alpha n_\mathrm{TL,DC}+\frac{1}{\tau_\mathrm{bulk}}+\frac{1}{\tau_\mathrm{tr}}\right)\\
        b&=\left(k_\mathrm{bulk}\alpha n_\mathrm{DC}-\frac{1}{\tau_\mathrm{tr}}\frac{n_\mathrm{TL,DC}\beta_\mathrm{DC}}{n_{\Phi}}-\frac{\beta_\mathrm{DC}}{\tau_\mathrm{tr}}\right)\\
        c&=-\left(\frac{1}{\tau_\mathrm{tr}\alpha}-k_\mathrm{int}n_\mathrm{TL,DC}\right) \\
        d&=\left(\frac{1}{\alpha\tau_\mathrm{tr}}\frac{n_\mathrm{TL,DC}\beta_\mathrm{DC}}{n_\Phi}+\frac{\beta_\mathrm{DC}}{\tau_\mathrm{tr}\alpha}+k_\mathrm{int}n_\mathrm{DC}+k_\mathrm{int}n_\mathrm{TL,DC}2\alpha\right)
    \end{align}
    By inversion of the matrix, the solution
    \begin{align}
        \tilde{n}&=\frac{i\omega + d}{\det(\mathbf{M})}\tilde{G}\\
        \tilde{n}_\mathrm{TL}&=\frac{-c}{\det(\mathbf{M})}\tilde{G}
    \end{align}
    can be obtained. 
    With 
    \begin{align}
        I_\mathrm{PL}&=k_\mathrm{rad}np
    \end{align}
    and neglecting all higher order and DC-terms, one arrives at
    \begin{align}
        \frac{\tilde{I}}{\tilde{G}}&=\frac{k_\mathrm{rad}}{\det(M)}\left(\left(2n_\mathrm{DC}+\alpha n_\mathrm{TL,DC}\right)\left(i\omega+d\right)-c\alpha n_\mathrm{DC}\right)\\
        &=\frac{k_\mathrm{rad}\left(2n_\mathrm{DC}+\alpha n_\mathrm{TL,DC}\right)}{\det(M)}\left(i\omega+d-\frac{c\alpha n_\mathrm{DC}}{\left(2n_\mathrm{DC}+\alpha n_\mathrm{TL,DC}\right)}\right)\\
        &=k_\mathrm{rad}\left(2n_\mathrm{DC}+\alpha n_\mathrm{TL,DC}\right)\frac{i\omega+\omega_{0}}{(i\omega+\omega_\mathrm{c1})(i\omega+\omega_\mathrm{c2})} \label{eq:SI_transfer_func}
    \end{align}
    where
    \begin{align}
        \det(\mathbf{M})&=(i\omega+\omega_\mathrm{c1})(i\omega+\omega_\mathrm{c2})
    \end{align}

    The three characteristic frequencies are
    \begin{align}
        \omega_{0}&=d-\frac{\alpha c n_\mathrm{DC}}{2n_\mathrm{DC}+\alpha n_\mathrm{TL,DC}} \label{eq:omega0}\\
        &=\left(\frac{1}{\alpha\tau_\mathrm{tr}}\frac{n_\mathrm{TL,DC}\beta_\mathrm{DC}}{n_\Phi}+\frac{\beta_\mathrm{DC}}{\alpha\tau_\mathrm{tr}}+2\alpha k_\mathrm{int}n_\mathrm{TL,DC}+n_\mathrm{DC}k_\mathrm{int}\right)+\frac{\alpha n_\mathrm{DC} \left(\frac{1}{\tau_\mathrm{tr}\alpha}-k_\mathrm{int}n_\mathrm{TL,DC}\right)}{2n_\mathrm{DC}+\alpha n_\mathrm{TL,DC}}\\
        \omega_\mathrm{c1,2}&=\frac{a+d}{2}\mp \sqrt{\left(\frac{a+d}{2}\right)^2-\left(ad-bc\right)}\\
        &=\frac{a+d}{2}\mp \frac{a+d}{2}\sqrt{1-\frac{ad-bc}{\left(\frac{a+d}{2}\right)^2}}\label{eq:omegac12}
    \end{align}
    where the root in the last expression can be Taylor expanded (assuming the two frequencies $\omega_\mathrm{c1}$ and $\omega_\mathrm{c2}$ are well separated, $\left(\frac{a+d}{2}\right)^2\gg \left(ad-bc\right)$) 
    \begin{align}    
        \omega_\mathrm{c1,2}&\approx\frac{a+d}{2}\mp \left(\frac{a+d}{2}\right)\left(1-\frac{ad-bc}{2\left(\frac{a+d}{2}\right)^2}\right)\\
        \omega_\mathrm{c1}&\approx\frac{ad-bc}{a+d}\\
        &=\frac{\left(k_\mathrm{bulk}2n_\mathrm{DC}+k_\mathrm{bulk}\alpha n_\mathrm{TL,DC}+\frac{1}{\tau_\mathrm{bulk}}+\frac{1}{\tau_\mathrm{tr}}\right)\left(\left(\frac{1}{\alpha\tau_\mathrm{tr}}\frac{n_\mathrm{TL,DC}\beta_\mathrm{DC}}{n_{\Phi}}+\frac{\beta_\mathrm{DC}}{\tau_\mathrm{tr}\alpha}+k_\mathrm{int}n_\mathrm{DC}+k_\mathrm{int}n_\mathrm{TL,DC}2\alpha\right)\right)}{\omega_\mathrm{c2}}\nonumber \\
        & ~~~~+\frac{\left(k_\mathrm{bulk}\alpha n_\mathrm{DC}-\frac{1}{\tau_\mathrm{tr}}\frac{n_\mathrm{TL,DC}\beta_\mathrm{DC}}{n_{\Phi}}-\frac{\beta_\mathrm{DC}}{\tau_\mathrm{tr}}\right)\left(\frac{1}{\tau_\mathrm{tr}\alpha}-k_\mathrm{int}n_\mathrm{TL,DC}\right)}{\omega_\mathrm{c2}} \\
        \omega_\mathrm{c2}&\approx a+d\\
        &=k_\mathrm{bulk}2n_\mathrm{DC}+k_\mathrm{bulk}\alpha n_\mathrm{TL,DC}+\frac{1}{\tau_\mathrm{bulk}}+\frac{1}{\tau_\mathrm{tr}}\\&~~~~+\left(\frac{1}{\alpha\tau_\mathrm{tr}}\frac{n_\mathrm{TL,DC}\beta_\mathrm{DC}}{n_{\Phi}}+\frac{\beta_\mathrm{DC}}{\tau_\mathrm{tr}\alpha}+k_\mathrm{int}n_\mathrm{DC}+k_\mathrm{int}n_\mathrm{TL,DC}2\alpha\right)
    \end{align}

\section{Additional Data of the Bilayers}
\label{sec:bilayer_data}
\begin{figure}[H]
        \centering
        \begin{subfigure}{0.32\textwidth}
            \centering
            \includegraphics[width=\linewidth]{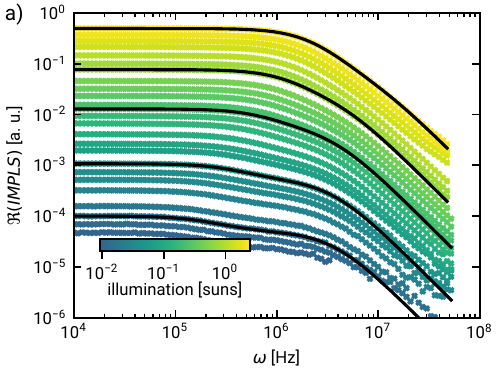}
        \end{subfigure}
        \hfill
        \begin{subfigure}{0.32\textwidth}
            \centering
            \includegraphics[width=\linewidth]{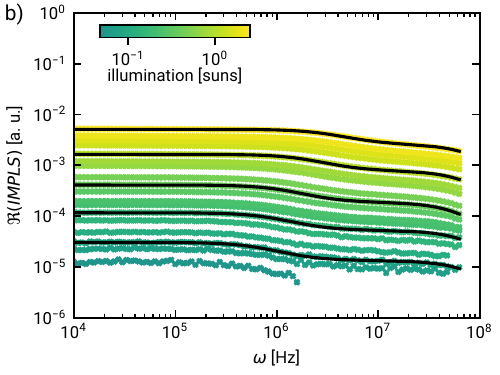}
        \end{subfigure}
        \hfill
        \begin{subfigure}{0.32\textwidth}
            \centering
            \includegraphics[width=\linewidth]{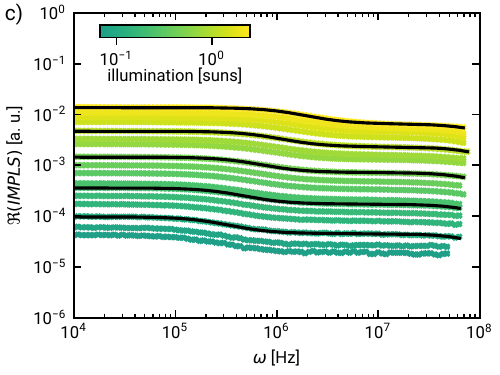}
        \end{subfigure}
        \caption{Real part of IMPLS data of the investigated perovskite/TL stacks: a)~SnO\textsubscript{2}/FAPbI\textsubscript{3}, b)~FAPbI\textsubscript{3}/spiro, c)~FAPbI\textsubscript{3}/OAI/spiro with exemplary fit lines.}
        \label{fig:real_bilayers}
    \end{figure}

\begin{figure}[H]
        \centering
        \begin{subfigure}{0.32\textwidth}
            \centering
            \includegraphics[width=\linewidth]{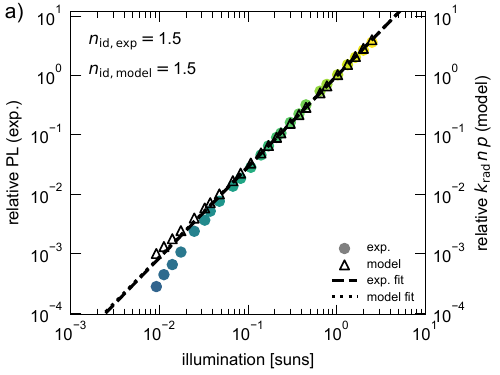}
        \end{subfigure}
        \hfill
        \begin{subfigure}{0.32\textwidth}
            \centering
            \includegraphics[width=\linewidth]{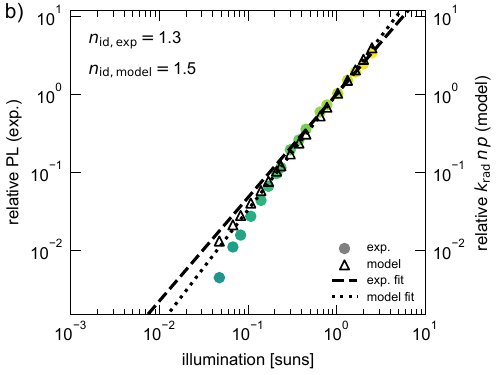}
        \end{subfigure}
        \hfill
        \begin{subfigure}{0.32\textwidth}
            \centering
            \includegraphics[width=\linewidth]{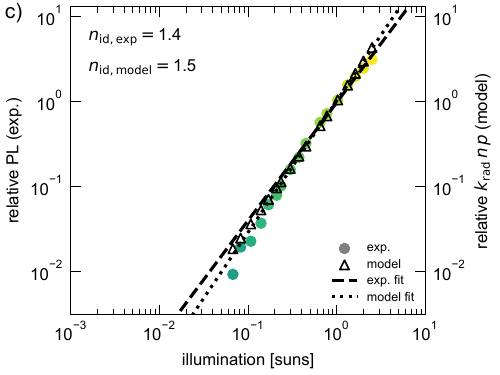}
        \end{subfigure}
        \caption{Suns-PL data of the investigated perovskite/TL stacks: a)~SnO\textsubscript{2}/FAPbI\textsubscript{3}, b)~FAPbI\textsubscript{3}/spiro, c)~FAPbI\textsubscript{3}/OAI/spiro with fit lines to intensities above 0.3 suns indicating the ideality factor in comparison to the relative PL intensity predicted from the zero-dimensional bilayer model.}
        \label{fig:suns_pl_bilayers}
    \end{figure}

\section{Fitting Procedure: Bulk/Transport Layer Model}
\label{sec:fitting}
A fit function was constructed consisting of two main parts: the first one determining the steady-state charge carrier concentrations by numerically solving~\cref{eq:dndt,eq:dntldt,eq:SI_charge_neutrality,eq:SI_beta} and the second one calculating the frequency domain response according to~\cref{eq:SI_transfer_func,eq:omega0,eq:omegac12}. This function was then fitted to the individually extracted characteristic frequencies and the relative steady-state PL intensity above 0.1 sun using a combination of differential evolution and least squares from scipy. Only data for $\omega> \qty{10}{\kilo\hertz}$ are fitted to minimize the influence of mobile ions. $T=\qty{300}{\kelvin}$ is assumed. The fit code along with sample data can be found at \url{https://github.com/deibellab/impls-bilayer-fit}.

\section{Uniqueness of Fit Results}
\label{sec:uniqe_fit}
\begin{figure}[H]
    \centering
    \includegraphics[width=0.8\linewidth]{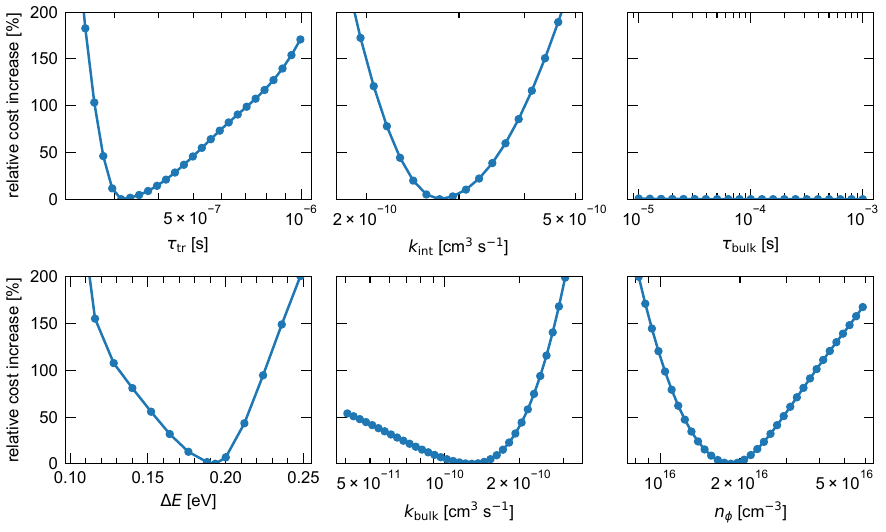}

    \caption{Relative cost increase of the bilayer model fit to the IMPLS response of the SnO\textsubscript{2}/FAPbI\textsubscript{3} stack when varying the parameter indicated on the x-axis and reoptimizing the others.}
    \label{fig:profile_sno2}
\end{figure}

\begin{figure}[H]
    \centering
    \includegraphics[width=0.8\linewidth]{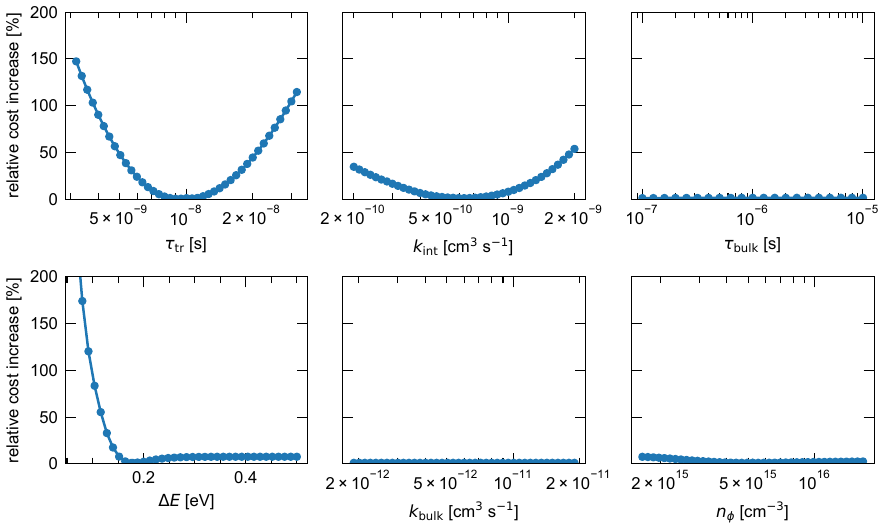}

    \caption{Relative cost increase of the bilayer model fit to the IMPLS response of the FAPbI\textsubscript{3}/spiro stack when varying the parameter indicated on the x-axis and reoptimizing the others.}
    \label{fig:profile_spiro}
\end{figure}

\begin{figure}[H]
    \centering
    \includegraphics[width=0.8\linewidth]{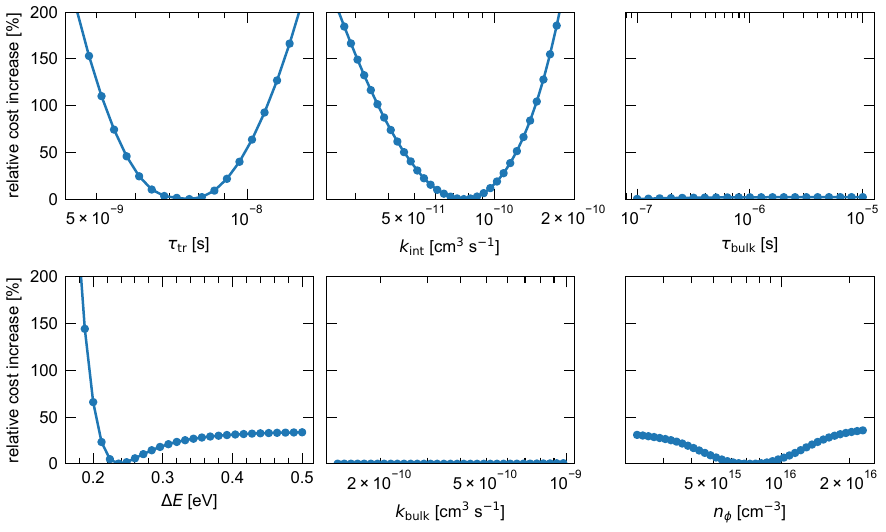}

    \caption{Relative cost increase of the bilayer model fit to the IMPLS response of the FAPbI\textsubscript{3}/OAI/spiro stack when varying the parameter indicated on the x-axis and reoptimizing the others.}
    \label{fig:profile_oai_spiro}
\end{figure}

To determine whether a certain parameter is actually constrained by the fit we apply the following technique for each parameter fit X. X is fixed to a certain value around its optimal value (fit result). Then the fit is reoptimized without adjusting X and the fit cost is recorded for different values of X. If other parameters can compensate for the change of X, the fit cost will not rise significantly. If a compensation is not possible the fit cost rises. \Cref{fig:profile_sno2,fig:profile_spiro,fig:profile_oai_spiro} show the relative cost increase of the global fits when varying the different parameters.

\section{Fitting the Bilayers Assuming Monomolecular Surface Recombination}
\label{sec:monomolecular_surface_rec}

    \begin{figure}[H]
        \centering
        \begin{subfigure}{0.32\textwidth}
            \centering
            \includegraphics[width=\linewidth]{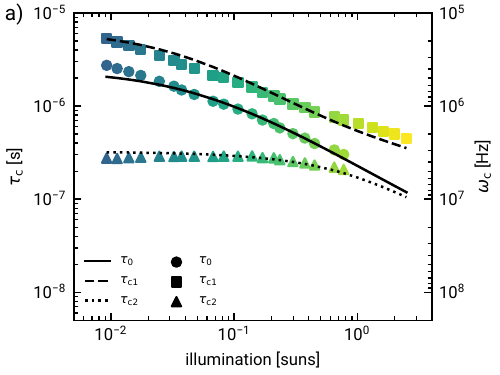}
        \end{subfigure}
        \hfill
        \begin{subfigure}{0.32\textwidth}
            \centering
            \includegraphics[width=\linewidth]{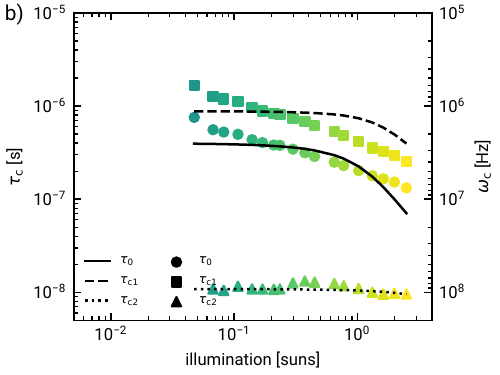}
        \end{subfigure}
        \hfill
        \begin{subfigure}{0.32\textwidth}
            \centering
            \includegraphics[width=\linewidth]{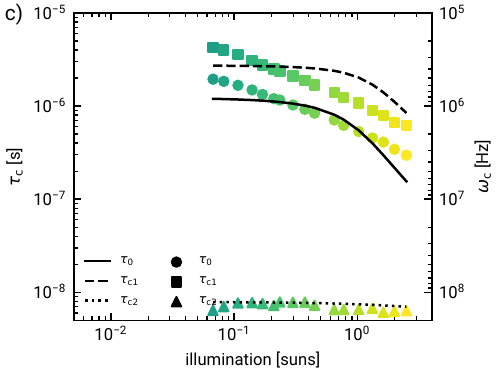}
        \end{subfigure}
        \caption{Model fits to the time constants of the investigated perovskite/TL stacks assuming monomolecular interface recombination: a)~SnO\textsubscript{2}/FAPbI\textsubscript{3}, b)~FAPbI\textsubscript{3}/spiro, c)~FAPbI\textsubscript{3}/OAI/spiro.}
        \label{fig:monoexp_surface_fits}
    \end{figure}

\section{Surface Recombination and Extraction Velocities}
\label{sec:surface_lifetime}

It is common to convert surface recombination and transfer rates into their respective velocities $S_\mathrm{int}$ and $S_\mathrm{tr}$, which we do through the relationships~\cite{Aalbers2026LowTemperatures,Kruckemeier2021UnderstandingTransient,Gillespie2024SiliconInspiredAnalysis,Sproul1994DimensionlessSolution,Wang2019ReducingSurface}
\begin{align}
    S_\mathrm{int}&=\frac{2 d}{\tau_\mathrm{int}-\frac{4 d^2}{D\pi^2}}\\
    S_\mathrm{tr}&=\frac{2d}{\tau_\mathrm{tr}-\frac{4 d^2}{D\pi^2}}\label{eq:s_transfer}
\end{align}
where $d$ is the thickness of the absorber layer and $D$ is the ambipolar diffusion constant (here assumed to be \qty{2.6e-1}{\centi\meter\squared\per\second} corresponding to a mobility of \qty{10}{\centi\meter\squared\per\volt\per\second}). The factor of two in the numerator is not used in all references, which might explain some of the deviations to reported values. The interface lifetime can be defined by multiplying \cref{eq:dntldt} by $\alpha$ and adding \cref{eq:dndt}
\begin{align}
    \frac{dp}{dt}&=G-k_\mathrm{bulk}np-\frac{n}{\tau_\mathrm{bulk}}
                   -\alpha k_\mathrm{int}n_\mathrm{TL}p
\end{align}
so the recombination rate of holes at the TL interface is
\begin{align}
    R_\mathrm{int}&=\alpha k_\mathrm{int}n_\mathrm{TL}p\\
    &=\frac{1}{\tau_\mathrm{int,p}}p
\end{align}
with $\tau_\mathrm{int,p}$ being the interface hole lifetime. In the case of an HTL the interface electron lifetime
\begin{align}
    R_\mathrm{int}&=\alpha k_\mathrm{int}p_\mathrm{TL}n\\
    &=\frac{1}{\tau_\mathrm{int,n}}n
\end{align}
can be found. Although $k_\mathrm{int}$ is well constrained by the fits, $\Delta E$ is not. As $n_\mathrm{TL}$ increases with increasing $\Delta E$, this leads to the surface lifetime and the respective surface recombination velocity not being very accurate.

For the SnO\textsubscript{2}/FAPbI\textsubscript{3} bilayer we derive a surface recombination velocity of \qty{80}{\centi\meter\per\second} at one sun and a transfer velocity of \qty{390}{\centi\meter\per\second}, both much lower than earlier reported values in the range of $S_\mathrm{int}=\qtyrange{250}{1000}{\centi\meter\per\second}$ and $S_\mathrm{tr}=\qty{5700}{\centi\meter\per\second}$~\cite{Gillespie2024SiliconInspiredAnalysis,Wang2019ReducingSurface,Brown2023DistinguishingElectron}. Studies do not always distinguish between the effects of transfer and recombination~\cite{Gillespie2024SiliconInspiredAnalysis,Wang2019ReducingSurface,Aalbers2025FunctionalizedSubstrates}, which might lead to the determination of an effective interface loss velocity $S_\mathrm{eff}=S_\mathrm{tr}+S_\mathrm{int}$~\cite{Aalbers2026LowTemperatures}. In our study, this results in \qty{470}{\centi\meter\per\second}, comparable to reported range of surface recombination velocities but not to the high reported transfer velocity.

As the transfer time is close to the diffusion term ($\approx\qty{6}{\nano\second}$) in \cref{eq:s_transfer}, the transfer velocity depends heavily on the assumed ambipolar diffusion coefficient or mobility. This might also explain (at least in part) the deviation from an earlier study, where an extraction velocity of only \qty{200}{\centi\meter\per\second} was reported~\cite{Butler-Caddle2024DistinguishingCarrier}, whereas our value is much larger: \qty{3e4}{\centi\meter\per\second}. Upon OAI-treatment, the across-interface recombination rate constant is reduced from \qty{6e-10}{\centi\meter\cubed\per\second} to \qty{8e-11}{\centi\meter\cubed\per\second}, which leads to a reduction of the electron surface recombination velocity at the interface from \qty{160}{\centi\meter\per\second} to \qty{50}{\centi\meter\per\second} at one sun each. The determined $S_\mathrm{int}$ of the untreated FAPbI\textsubscript{3}/spiro interface is well below reported values in the range of thousands of \unit{\centi\meter\per\second}~\cite{Gillespie2024SiliconInspiredAnalysis,Wang2019ReducingSurface}.

\section{Fit of the Kinetic Model without the Electrostatic Term}
\label{sec:fit_no_phi}
\begin{figure}[H]
        \centering
        \includegraphics{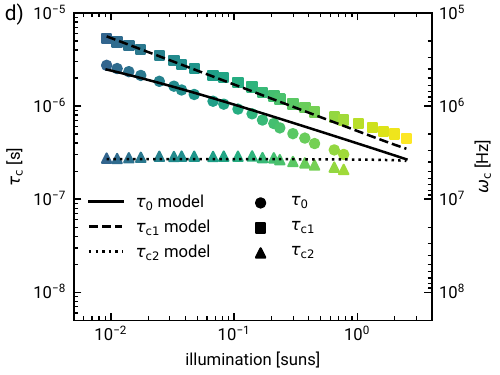}
        \caption{Model fit to the time constants of the investigated SnO\textsubscript{2}/FAPbI\textsubscript{3} bilayer without the electrostatic term.}
        \label{fig:fit_no_electrostatics}
    \end{figure}

\section{Equation Set for Predicting an Upper Limit of the Device Lifetime}
\label{sec:upper_limit}

Rate equations including an explicit population of the HTL:
\begin{align}
    \frac{dn}{dt}&=G-\frac{1}{\tau_\mathrm{tr,ETL}}\left(n-n_\mathrm{ETL}\beta_\mathrm{ETL}\right)-\alpha_\mathrm{HTL} k_\mathrm{int,HTL}p_\mathrm{HTL}n\\
    \frac{dn_\mathrm{ETL}}{dt}&=\frac{1}{\tau_\mathrm{tr,ETL}\alpha_\mathrm{ETL}}\left(n-n_\mathrm{ETL}\beta_\mathrm{ETL}\right)- k_\mathrm{int,ETL}n_\mathrm{ETL}p\\
    \frac{dp_\mathrm{HTL}}{dt}&=\frac{1}{\tau_\mathrm{tr,HTL}\alpha_\mathrm{HTL}}\left(p-p_\mathrm{HTL}\beta_\mathrm{HTL}\right)- k_\mathrm{int,HTL}p_\mathrm{HTL}n\\
    p &= \alpha_\mathrm{ETL} n_\mathrm{ETL}+n -\alpha_\mathrm{HTL} p_\mathrm{HTL}
\end{align}
with $\alpha_\mathrm{ETL}=d_\mathrm{ETL}/d$, $\alpha_\mathrm{HTL}=d_\mathrm{HTL}/d$ and
\begin{align}
    \beta_\mathrm{ETL}&=\exp\left(-\frac{\Delta E_\mathrm{ETL}}{k_\mathrm{B}T}
                        +\frac{n_\mathrm{ETL}}{n_{\Phi,\mathrm{ETL}}}\right)\\
    \beta_\mathrm{HTL}&=\exp\left(-\frac{\Delta E_\mathrm{HTL}}{k_\mathrm{B}T}
                        +\frac{p_\mathrm{HTL}}{p_{\Phi,\mathrm{HTL}}}\right)
\end{align}
$\frac{dp}{dt}$ is redundant when enforcing charge neutrality. $p_\Phi$ is the analogue of $n_\Phi$ for the HTL. When numerically solving the system, all parameters with HTL as a subscript are taken from the fit results of the spiro bilayers (without OAI for the control cell and with OAI for the treated cell); all parameters with ETL as a subscript are taken from the fit results of the bilayer with SnO\textsubscript{2}. Bulk recombination is neglected in this case and $T=\qty{300}{\kelvin}$.
The effective lifetime is calculated by
\begin{align}
    \tau_\mathrm{rec}=\frac{n+p}{2G}
\end{align}
For the case of a thin film with $n=p$ this reduces to \cref{eq:simple_tau}.

\section{Derivation of the Transfer Functions for IMVS and IMPLS}
\label{sec:derivation_imvs_impls}
Starting from eq. (25) in~\cite{Ravishankar2023HowCharge} we can show the general shape of the transfer function of IMVS.

\begin{align}
\frac{\tilde{V}_\mathrm{elec}}{\tilde{j}_{\Phi}}&=\left(\frac{1}{R_\mathrm{rec}} + i\omega C_{\mu} +i\omega C_\mathrm{g} \left(1+\frac{R_\mathrm{exc}}{R_\mathrm{rec}} + i\omega R_\mathrm{exc}C_{\mu}\right)\right)^{-1}\\
&=\left(\frac{1}{R_\mathrm{rec}} + i\omega \left(C_{\mu}+C_\mathrm{g}+C_\mathrm{g}\frac{R_\mathrm{exc}}{R_\mathrm{rec}}\right)+\left(i\omega\right)^2C_\mathrm{g}C_{\mu}R_\mathrm{exc} \right)^{-1}\label{eq:transfer_func_imvs}\\
A&=C_\mathrm{g}C_{\mu}R_\mathrm{exc}\label{eq:A}\\
B&=C_{\mu}+C_\mathrm{g}+C_\mathrm{g}\frac{R_\mathrm{exc}}{R_\mathrm{rec}}\label{eq:B}\\
C&=\frac{1}{R_\mathrm{rec}}\label{eq:C}\\
\omega_\mathrm{c1,2}&=\frac{B\pm\sqrt{B^2-4AC}}{2A}\\
\frac{\tilde{V}_\mathrm{elec}}{\tilde{j}_{\Phi}}&=\frac{1}{C_\mathrm{g}C_{\mu}R_\mathrm{exc}\left(i\omega+\omega_\mathrm{c1}\right)\left(i\omega+\omega_\mathrm{c2}\right)}
\end{align}

Starting from eq. (22) in~\cite{Ravishankar2023HowCharge} we can derive a transfer function for IMPLS $\frac{\tilde{V}_\mathrm{int}}{\tilde{j}_{\Phi}}$
\begin{align}
    i\omega C_{\mu}\tilde{V}_\mathrm{int}&=-\left(\frac{1}{R_\mathrm{exc}} + \frac{1}{R_\mathrm{rec}}\right)\tilde{V}_\mathrm{int}+\frac{1}{R_\mathrm{exc}}\tilde{V}_\mathrm{elec}+\tilde{j}_{\Phi}\label{eq:ravishankar1}\\
    i\omega C_\mathrm{g}\tilde{V}_\mathrm{elec}&=\frac{1}{R_\mathrm{exc}}\tilde{V}_\mathrm{int}-\frac{1}{R_\mathrm{exc}}\tilde{V}_\mathrm{elec} \label{eq:ravishankar2}
\end{align}
Solving \cref{eq:ravishankar2} for $\tilde{V}_\mathrm{elec}$
\begin{align}
    \tilde{V}_\mathrm{elec}&=\frac{\tilde{V}_\mathrm{int}}{\left(i\omega C_\mathrm{g}R_\mathrm{exc}+1\right)} \label{eq:V_elec}
\end{align}
Plug \cref{eq:V_elec} into \cref{eq:ravishankar1}
\begin{align}
    \frac{\tilde{V}_\mathrm{int}}{\tilde{j}_{\Phi}}&=\left(i\omega C_{\mu} + \frac{1}{R_\mathrm{rec}}+\frac{i\omega C_\mathrm{g}}{\left(i\omega C_\mathrm{g}R_\mathrm{exc}+1\right)}\right)^{-1}\\
    &=\frac{\left(i\omega C_\mathrm{g}R_\mathrm{exc}+1\right)}{\left(i\omega\right)^{2}C_{\mu}C_\mathrm{g}R_\mathrm{exc}+i\omega\left(C_\mu+C_\mathrm{g}\frac{R_\mathrm{exc}}{R_\mathrm{rec}}+C_\mathrm{g}\right)+\frac{1}{R_\mathrm{rec}}}\\
    &\propto \frac{i\omega+\omega_0}{\left(i\omega+\omega_\mathrm{c1}\right)\left(i\omega+\omega_\mathrm{c2}\right)}
\end{align}
The denominator is identical to the one in \cref{eq:transfer_func_imvs}, the numerator gains an extra characteristic frequency
\begin{align}
    \omega_0=\frac{1}{C_\mathrm{g}R_\mathrm{exc}}=\frac{1}{\tau_\mathrm{elec}} \label{eq:omega0B}
\end{align}

From \cref{eq:transfer_func_imvs,eq:A,eq:B,eq:C,eq:omega0B} the relationships between the physical time scales and the measured characteristic frequencies can be derived
\begin{align}
    \omega_\mathrm{c1}+\omega_\mathrm{c2}&=\frac{B}{A}\\
    \omega_\mathrm{c1}\omega_\mathrm{c2}&=\frac{C}{A}\\
    \tau_\mathrm{eff}&=\frac{\omega_0}{\omega_\mathrm{c1}\omega_\mathrm{c2}}\\
    \tau_\mathrm{exc}&=\frac{1}{\omega_\mathrm{c1}+\omega_\mathrm{c2}-\omega_0-\frac{\omega_\mathrm{c1}\omega_\mathrm{c2}}{\omega_0}}
\end{align}

\section{IMPLS and IMVS of the Control Cell}
\label{sec:data_control_cell}

    \begin{figure}[H]
        \centering

        \begin{subfigure}{0.48\textwidth}
            \centering
            \includegraphics{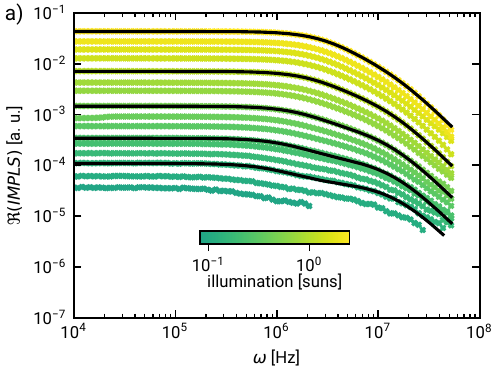}
        \end{subfigure}
        \hfill
        \begin{subfigure}{0.48\textwidth}
            \centering
            \includegraphics{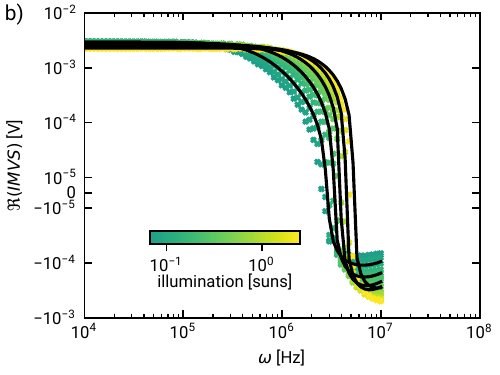}
        \end{subfigure}

        \vspace{0.5em}

        \begin{subfigure}{0.48\textwidth}
            \centering
            \includegraphics{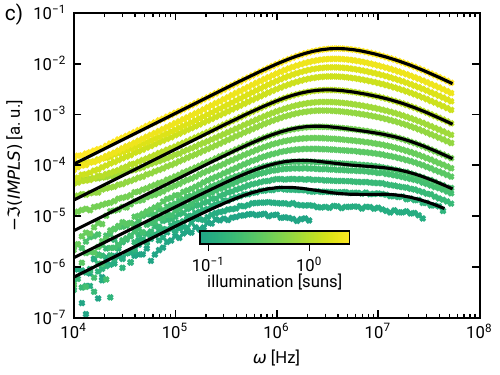}
        \end{subfigure}
        \hfill
        \begin{subfigure}{0.48\textwidth}
            \centering
            \includegraphics{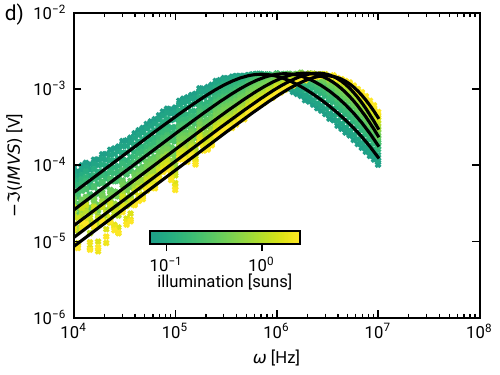}
        \end{subfigure}

        \caption{Real (top) and imaginary (bottom) part of IMPLS (left) and IMVS (right) for the control cell.}
        \label{fig:control_ims}
    \end{figure}

    \begin{figure}[H]
        \centering
        \begin{subfigure}{0.48\textwidth}
            \centering
            \includegraphics{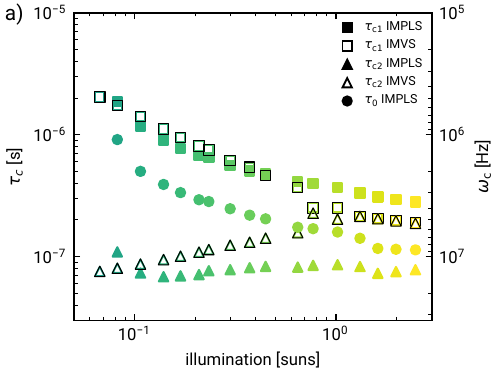}
        \end{subfigure}
        \hfill
        \begin{subfigure}{0.48\textwidth}
            \centering
            \includegraphics{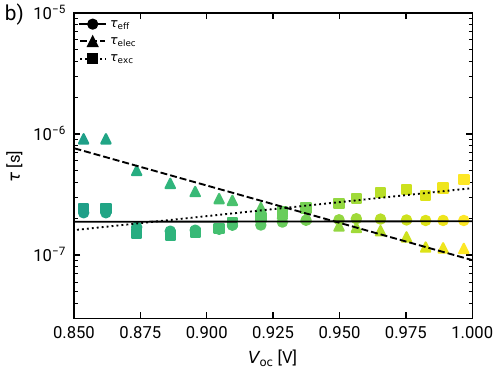}
        \end{subfigure}
        \caption{a) Extracted time constants/ characteristic frequencies from IMPLS and IMVS for different light intensities, b) corresponding physical time scales.}
        \label{fig:time_constants_control}
    \end{figure}

\section{Additional Data of the OAI-Treated Solar Cells}
\label{sec:data_oai_cell}
    \begin{figure}[H]
        \centering
        \begin{subfigure}{0.48\textwidth}
            \centering
            \includegraphics{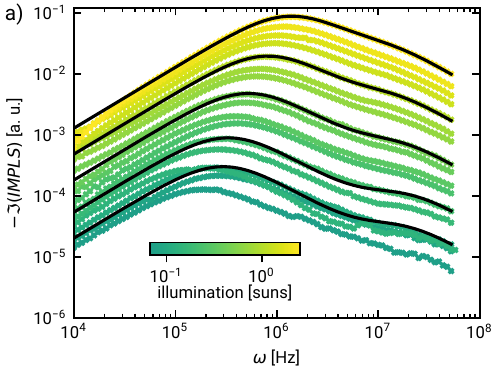}
        \end{subfigure}
        \hfill
        \begin{subfigure}{0.48\textwidth}
            \centering
            \includegraphics{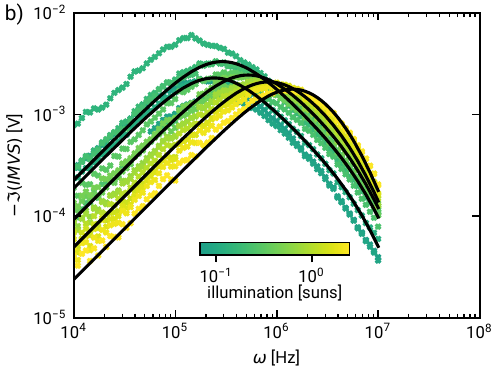}
        \end{subfigure}
        \caption{Imaginary part of a) IMPLS and b) IMVS of the OAI-treated cell.}
        \label{fig:imag_cell}
    \end{figure}

\section{Ideality Factor from Open-Circuit Voltage Measurements}
\label{sec:ideality_voc}
    In order to determine the ideality factor from electrical measurements we determine the relative steady-state $V_\mathrm{OC}$ with varying illumination intensity.  The ideality factor is calculated by~\cite{Kirchartz2012MeaningReaction,Calado2019IdentifyingDominant,Caprioglio2019RelationOpenCircuit}
    \begin{align}
        n_\mathrm{id,Voc} = \frac{q}{k_\mathrm{B}T}\frac{d V_\mathrm{OC}}{d \ln\left(G\right)}
    \end{align}
    
    \begin{figure}[H]
        \centering
        \begin{subfigure}{0.48\textwidth}
            \centering
            \includegraphics{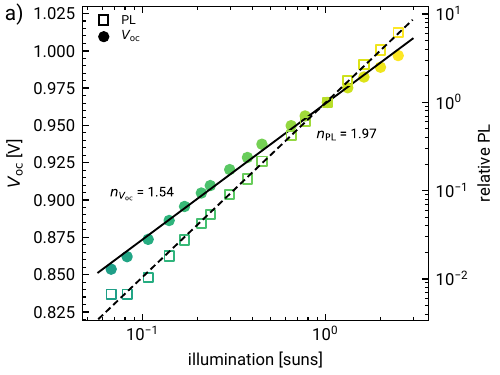}
        \end{subfigure}
        \hfill
        \begin{subfigure}{0.48\textwidth}
            \centering
            \includegraphics{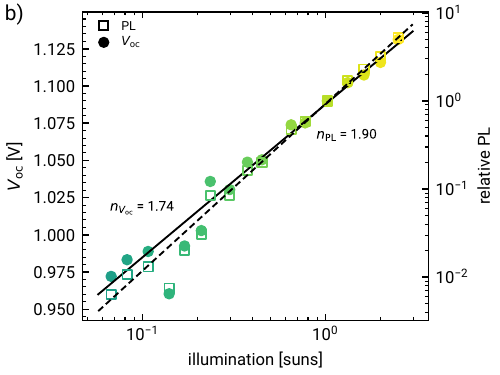}
        \end{subfigure}
        \caption{Suns-$V_\mathrm{OC}$ and suns-PL data of the investigated solar cells: a) control device, b) OAI-treated device with fit lines indicating the ideality factors.}
        \label{fig:suns-voc-pl}
    \end{figure}

\section{Charge Carrier Exchange}
\label{sec:charge_carrier_exchange}

The charge carrier exchange time increases from below \qty{100}{\nano\second} to about \qty{200}{\nano\second} within the studied illumination window. This slowing down of charge exchange between bulk and electrodes is expected as the electric field in the transport layer reduces at higher $V_\mathrm{OC}$s~\cite{Ravishankar2024DiscerningRise}. At one sun, the charge carrier exchange time of $\qty{160}{\nano\second}$ corresponds to a charge carrier exchange velocity $S_\mathrm{exc}=d/\tau_\mathrm{exc}$~\cite{Ravishankar2023HowCharge} of $\qty{380}{\centi\meter\per\second}$. As the concept of this exchange velocity is not yet widely adopted, literature values for comparison are sparse. Our calculated value lies above the range of \qtyrange{1}{200}{\centi\meter\per\second} for Poly[bis(4-phenyl)(2,4,6-trimethylphenyl)amine] (PTAA)-based devices~\cite{Kruckemeier2023QuantifyingCharge,Ravishankar2023HowCharge}, which is consistent with the spiro and the SnO\textsubscript{2} having a higher mobility than the PTAA. A value derived from voltage-dependent PL on a device with a self-assembled monolayer and C60 as TLs ($S_\mathrm{exc}\approx \qty{2700}{\centi\meter\per\second}$)~\cite{Yuan2025DerivingMobilityLifetime} exceeds our value by almost an order of magnitude.

The data shows $\tau_\mathrm{elec}$ decreasing with light intensity. Assuming $C_\mathrm{g}$ to be constant, looking at eq. (11) in the main text and the definition of the exchange resistance $R_\mathrm{exc}=d/\left(C_\mu S_\mathrm{exc}\right)$ this means that the exchange velocity decreases less than the chemical capacitance increases.

\section{Materials, Device Fabrication and Experimental Methods}
\textbf{Materials}

Fluorine-doped tin oxide on glass (FTO glass, \qty{8}{\ohm\per\sq}) was purchased from Pilkington. Tin(IV) oxide (15 wt\% in H\textsubscript{2}O colloidal dispersion) was purchased from Alfa Aesar. Lead iodide (99.99\%) was purchased from TCI Chemicals. Formamidinium iodide (FAI, 99.99\%), methylammonium chloride (MACl, 99.99\%), \textit{n}-octylammonium iodide (OAI, 99\%), and FK209 Co(III) TFSI salt (FK209) were purchased from Greatcell Solar Materials. Deionized water, potassium chloride (KCl, $\ge$99\%), cesium chloride (CsCl, 99.999\%), lithium bis(trifluoromethane)sulfonimide (Li-TFSI, 99.95\%), spiro-OMeTAD (99\%), and 4-\textit{tert}-butylpyridine (4-\textit{t}BP, 98\%) were purchased from Sigma-Aldrich. \textit{N,N}-dimethylformamide (DMF, 99.8\%, ExtraDry), dimethyl sulfoxide (DMSO, 99.9\%), chlorobenzene (CB, 99.8\%, ExtraDry), isopropanol (IPA, 99.8\%, ExtraDry), and acetonitrile (ACN, 99.9\%) were purchased from Acros Organics. All materials were used as received without further purification.

\noindent\textbf{FAPbI\textsubscript{3} Synthesis}

PbI\textsubscript{2} (\qty{4.61}{\gram}, \qty{10}{\milli\mol}) and FAI (\qty{1.72}{\gram}, \qty{10}{\milli\mol}) were dissolved in \qty{10}{\milli\liter} of 2-methoxyethanol and reacted at \qty{120}{\celsius} for \qty{2}{\hour}. The resulting black powder was collected by filtration and washed with acetonitrile and subsequently diethyl ether. The obtained FAPbI\textsubscript{3} powder was heated at \qty{150}{\celsius} for \qty{30}{\minute} on a hot plate and dried overnight in a vacuum oven at \qty{50}{\celsius}. 

\noindent\textbf{Device Fabrication}

FTO glass substrates were sequentially cleaned in an ultrasonic bath of deionized water, acetone, and isopropanol for \qty{10}{\minute} each. The SnO\textsubscript{2} colloidal dispersion was diluted with DI water at a volume ratio of 1:4 (SnO\textsubscript{2} dispersion:DI water). After UV-ozone treatment of the FTO substrates for \qty{20}{\minute}, the diluted SnO\textsubscript{2} solution was spin-coated on the FTO substrates at \qty{5000}{rpm} for \qty{60}{\second} and then annealed at \qty{170}{\celsius} for \qty{30}{\minute}. \qty{50}{\micro\liter} of KCl solution (\qty{3}{\milli\gram\per\milli\liter} in DI water) was dropped onto the substrate and spin-coated at \qty{4000}{rpm} for \qty{40}{\second}, followed by annealing at \qty{100}{\celsius} for \qty{10}{\minute}. For the perovskite layer, FAPbI\textsubscript{3} powder (\qty{633}{\milli\gram}), MACl (\qty{13.5}{\milli\gram}), and CsCl (\qty{8.4}{\milli\gram}) were dissolved in a mixed solvent of DMF (\qty{560}{\micro\liter}) and DMSO (\qty{74.4}{\micro\liter}). The precursor solution was spin-coated at \qty{4000}{rpm} for \qty{30}{\second}, and diethyl ether (\qty{400}{\micro\liter}) was dropped on the rotating substrate \qty{10}{\second} before the end of the spin-coating process. The substrates were annealed at \qty{150}{\celsius} for \qty{15}{\minute}. \qty{50}{\micro\liter} of OAI solution (\qty{15}{mM} in IPA) was spin-coated at \qty{5000}{rpm} for \qty{20}{\second}. The spiro-OMeTAD solution (\qty{90}{\milli\gram\per\milli\liter} in CB) was doped with \qty{39}{\micro\liter} of 4-\textit{t}BP, \qty{23}{\micro\liter} of Li-TFSI (\qty{520}{\milli\gram\per\milli\liter} in ACN), and \qty{5}{\micro\liter} of FK209 (\qty{325}{\milli\gram\per\milli\liter} in ACN), followed by spin-coating at \qty{5000}{rpm} for \qty{20}{\second}. Finally, \qty{100}{\nano\meter} of Au was thermally evaporated through a shadow mask, defining a device area of \qty{0.045}{\centi\meter\squared}. 

\noindent\textbf{J-V Measurements}

The $j$--$V$ characteristics of the solar cells were measured using a Keithley 2450 source meter unit. The photovoltaic performance was evaluated under simulated AM 1.5G illumination at an intensity of \qty{100}{\milli\watt\per\centi\meter\squared} using an Abet Sun 3000 Class AAA solar simulator. The light intensity was calibrated using a Si reference cell (VLSI Standards Inc.). The spectral mismatch factor was determined by considering the AM 1.5G reference spectrum and the spectral responses of the Si reference cell and perovskite devices. All devices were scanned from \qty{1.3}{\volt} to \qty{-0.3}{\volt} at a scan rate of \qty{0.02}{\volt\per\second}. The active area was \qty{0.045}{\centi\meter\squared}, and the measurements were performed using an aperture under ambient conditions.  

\noindent\textbf{IMPLS on Thin Films and Bilayers}

IMPLS measurements were conducted with the sample placed in nitrogen atmosphere inside a Cryovac cryostat. A Zurich Instruments HF2LI lock-in amplifier was used to supply the modulation and DC bias to the laser (PhoxX+, \qty{515}{\nano\meter}, Omikron). The light intensity was set by a pair of filter wheels (Thorlabs) keeping the AC/DC ratio at \qty{10}{\percent} throughout all light intensities. The laser was coupled into a multimode fiber at a non-perpendicular angle resulting in a homogeneous spot at the output. Spot size on the sample was approximately \qty{4}{\milli\meter} in diameter. PL was collected by a lens system and measured using an avalanche photodiode (APD130, Thorlabs) with two long pass filters blocking scattered laser light. A second avalanche photodiode was used to measure the laser reference simultaneously. The lock-in was used to acquire the signal from both detectors. The IMPLS response was corrected for the frequency dependent phase shift and amplitude of the system. The frequency range was f=\qty{1}{\hertz} to \qty{50}{\mega\hertz}. Illumination level was defined by measuring the short circuit current density of a corresponding solar cell.

\noindent\textbf{IMVS and IMPLS on Solar Cells}

A metal mask was placed on the sample ensuring only PL from one pixel was collected. The setup was identical to the one for IMPLS on thin films and bilayers, except connecting the solar cell to the lock-in amplifier instead of the reference detector. IMVS and IMPLS were measured simultaneously. Reference measurements were carried out afterwards. The IMVS reference was measured with a fast photodiode (DET025A, Thorlabs).

\noindent\textbf{PLQY}

PLQY measurements were carried out in an integrating sphere (Ocean Optics) in ambient atmosphere. A laser (PhoxX+, \qty{515}{\nano\meter}, Omikron) in conjunction with a multimode fiber under non-perpendicular coupling provided uniform illumination. Laser intensity was set to match the one sun equivalent photon flux.

\noindent\textbf{EQE}

Relative EQE measurements were conducted using a Bentham light source and lock-in amplifier with a chopper wheel and a silicon photodiode (FDS1010, Thorlabs) serving as a reference. EQE was normalized by the measured short circuit current density.

\clearpage
\printbibliography

\end{refsection}

\end{document}